\documentclass{aa}

\usepackage{graphicx}
\graphicspath{{figures}}
\usepackage{newtxtext, newtxmath}
\usepackage{placeins}           

\usepackage{hyperref}
\hypersetup{breaklinks=true, colorlinks=true, allcolors=blue}

\usepackage{microtype}
\usepackage{orcidlink}
\usepackage{siunitx}
\DeclareSymbolFont{upgreek}{U}{eur}{m}{n}
\DeclareMathSymbol{\umu}{0}{upgreek}{"16}
\DeclareSIPrefix{\micro}{\text{\ensuremath{\umu}}}{-6}

\let\oldbibliography\thebibliography
\renewcommand{\thebibliography}[1]{%
    \oldbibliography{#1}%
    \setlength{\itemsep}{0pt}%
}

\begin{document}

\title{%
    Polarization at millimeter wavelengths caused by drifting grains\\ in protoplanetary disks
}


\titlerunning{%
    Polarization caused by drifting grains in protoplanetary disks
}

\author{%
    Moritz Lietzow-Sinjen\inst{1}\fnmsep\thanks{Corresponding author; \href{mailto:mlietzow@astrophysik.uni-kiel.de}{\tt mlietzow@astrophysik.uni-kiel.de}}\,\orcidlink{0000-0001-9511-3371}
    \and
    Stefan Reissl\inst{2}\,\orcidlink{0000-0001-5222-9139}
    \and
    Mario Flock\inst{3}\,\orcidlink{0000-0002-9298-3029}
    \and
    Sebastian Wolf\inst{1}\,\orcidlink{0000-0001-7841-3452}
}

\authorrunning{%
    Lietzow-Sinjen~\etalname
}

\institute{%
    Institut für Theoretische Physik und Astrophysik, Christian-Albrechts-Universität zu Kiel, Leibnizstr.~15, 24118 Kiel, Germany
    \and
    Zentrum für Astronomie der Universität Heidelberg, Institut für Theoretische Astrophysik, Albert-Ueberle-Str.~2, 69120 Heidelberg, Germany
    \and
    Max-Planck-Institut für Astronomie, Königstuhl~17, 69117 Heidelberg, Germany
}

\date{%
    Received 24 December 2024 / Accepted 02 September 2025
}

\abstract
{
    During the evolution of protoplanetary disks, dust grains start to grow, form larger particles, settle to the midplane, and rearrange the disk, mainly by the inward radial drift.
    Because of this, dust pebbles with an irregular shape usually align mechanically and thus cause polarization signatures in their thermal radiation due to dichroic emission or absorption.
}
{
    The goal of this paper is to evaluate the potential to trace the impact of mechanical grain alignment in protoplanetary disks on the observed degree and orientation of linear polarization at millimeter wavelengths.
}
{
    We combined 3D radiation hydrodynamical simulations to determine the density distribution and the velocity field of gas and dust particles, Monte Carlo dust-gas interaction simulations to calculate the mechanical alignment of dust in a gas flow, and, finally, 3D Monte Carlo polarized radiative transfer simulations to obtain synthetic polarimetric observations.
}
{
    We find that large grains, which contribute the most to the net polarization, are potentially mechanically aligned in the protoplanetary disk under the effect of the vertical shear instability (VSI).
    Thereby, the drift velocity is parallel to the rotational disk axis.
    Assuming oblate dust grains that are aligned with their short axis parallel to the direction of the drift velocity, the resulting polarization is usually along the major axis of the disk.
    This is in contrast to typical drift models that propose either a radial or azimuthal drift velocity component.
}
{
    If hydrodynamical instabilities, such as the VSI, dominate the kinematics in protoplanetary disks, the mechanical alignment of dust is a promising mechanism for grain alignment in these systems.
    In that case, the resulting millimeter polarization allows us to trace the orientation of aligned millimeter-sized grains.
}


\maketitle
\nolinenumbers


\section{Introduction}
\label{sec:introduction}

Recent high-resolution observations of young stellar objects with the Atacama Large Millimeter/submillimeter Array (ALMA) at (sub)millimeter wavelengths reveal several substructures in their circumstellar disks, such as rings, spirals, or gaps \citep[e.g.,][]{ALMA_Partnership_etal_2015, Andrews_etal_2018}.
These substructures are either attributed to still evolving and forming but yet unseen giant planets or brown dwarfs \citep{Brown_etal_2009, Andrews_etal_2011, Biller_etal_2012}, or to magneto-hydrodynamical instabilities that play an important role during the accretion process.

In the inner disk regions, that is, inside \SI{1}{au}, the magneto-rotational instability \citep{Balbus_Hawley_1991, Flock_etal_2011, Turner_etal_2014} and magnetically driven winds \citep{Bai_etal_2016, Gressel_etal_2020, Lesur_2021} determine the accretion flow and shape of the disk structure \citep{Flock_etal_2017a, Lesur_etal_2023}.
In contrast, for less ionized regions in the outer disk, hydrodynamical instabilities might become more important in determining the gas and dust flows.
Most notably, the vertical shear instability \citep[VSI,][]{Urpin_Brandenburg_1998, Urpin_2003, Arlt_Urpin_2004, Nelson_etal_2013, Stoll_Kley_2014, Lin_Youdin_2015, Flock_etal_2017b, Schaefer_etal_2020, Flores-Rivera_etal_2020, Barraza-Alfaro_etal_2021}, which is related to the Goldreich-Schubert-Fricke instability \citep{Goldreich_Schubert_1967, Fricke_1968}, is a promising mechanism that causes turbulence in the regions outside \SI{10}{au} of protoplanetary disks.
This instability arises from fluctuations in the orbital motion in the vertical direction, that is, along the rotational axis of the disk.

Additionally, it has been shown that the VSI can cause the formation of vortices, which act as particle traps and are therefore essential in the context of planet formation \citep{Richard_etal_2016, Latter_Papaloizou_2018, Manger_Klahr_2018}.
In this work, we use simulation results from high-resolution 3D radiation hydrodynamical simulations of the outer regions of protoplanetary disks presented in \citet{Flock_etal_2020}.
This work also includes the individual motion of larger grains as a result of the gas drag.

Once interstellar micrometer-sized grains reach higher densities in the disk, they start to grow and settle to the midplane of the disk \citep{Beckwith_etal_2000}.
Micrometer-sized grains are still well coupled to the gas motion.
However, once they grow to sizes of hundreds of micrometers, their motion starts to decouple from that of the gas phase \citep{Weidenschilling_1977}.
During this phase of dust evolution, a large amount of solid material, in the form of millimeter- and centimeter-sized pebbles, is being rearranged in the young protoplanetary disk, mainly by the fast inward radial drift and the vertical migration toward the disk midplane \citep{Birnstiel_etal_2010, Andrews_2020, Birnstiel_2024}.

These irregularly shaped dust pebbles typically align on large scales along a preferential direction relative to a given axis due to torques acting upon them.
Such torques can arise from radiation, magnetic fields, or mechanical gas drag.
The most promising alignment mechanism is radiative torque (RAT) alignment \citep[e.g.,][]{Dolginov_Mitrofanov_1976, Draine_Weingartner_1996, Draine_Weingartner_1997, Lazarian_Hoang_2007a, Lazarian_Hoang_2007b, Hoang_Lazarian_2008}.
In this scenario, radiation spins up the dust grains, and the rotation axis subsequently aligns with the magnetic field orientation due to paramagnetic dissipation, a mechanism that has been extensively studied in the context of circumstellar disks \citep{Cho_Lazarian_2007, Reissl_etal_2016, Tazaki_etal_2017}.
As a result, the net thermally re-emitted radiation from the aligned, elongated grains is expected to be partially polarized, with the observable polarization vectors tracing the magnetic field orientation \citep[see, e.g.,][for a review]{Andersson_2015}.

However, the question of which mechanisms, other than RATs, can drive coherent grain alignment remains an active area of research \citep{Dolginov_Mitrofanov_1976, Purcell_1979, Andersson_2015, Hoang_etal_2018}.
A mechanism based on supersonic gas-dust drift was proposed more than half a century ago \citep[][]{Gold_1952a, Gold_1952b}.
Such a gas-dust drift would mechanically align spheroidal grains by minimizing their geometrical cross section.
In this scenario, the resulting dust polarization is related to the direction of the drift velocity and does not trace the magnetic field.
A description of mechanical alignment based on a toy model was provided by \citet{Lazarian_Hoang_2007a}, \citet{Das_Weingartner_2016}, and \citet{Reissl_etal_2023}, suggesting that irregularly shaped dust grains may also attain a long-term stable alignment configuration even under subsonic drift conditions.
In the case of mechanical alignment, the helicity of an oblate grain can be either right-handed or left-handed \citep{Kataoka_etal_2019}.
In both cases, the resulting polarization is oriented perpendicular to the direction of the gas velocity.

Consequently, measuring the degree of polarization of dust thermal emission is a potentially powerful tool for constraining the physical processes related to gas flow in protoplanetary disks.
Therefore, in this study, we do not attempt to reproduce previous efforts to model dust polarization in disks based on RATs and magnetic field alignment \citep[e.g.,][]{Cho_Lazarian_2007,Tazaki_etal_2017}, but instead focus on the processes that enable the mechanical alignment of dust pebbles.
Specifically, the goal of this study is to analyze the impact of mechanical grain alignment caused by the drift motion of these pebbles in the protoplanetary disk on the resulting observable linear polarization.

We used simulation snapshots from 3D radiation hydrodynamical simulations, including the motion of larger dust pebbles \citep{Flock_etal_2020}.
Subsequently, we constrained the characteristics of wavelength-dependent polarization maps resulting from the polarized emission of aligned grains.
Here, we consider synthetic observations in the millimeter wavelength range, that is, from \SI{0.87}{mm} to \SI{10}{mm}, to trace the thermal emission of large millimeter-sized dust grains.

This paper is organized as follows.
In Sect.~\ref{sec:model_setup}, we introduce and describe our model setup and numerical procedure.
Subsequently, we present and discuss synthetic observations generated using radiative transfer simulations in Sect.~\ref{sec:results}.
In Sect.~\ref{sec:limitations}, we discuss several assumptions and limitations in our grain alignment model, and finally we summarize our study in Sect.~\ref{sec:conclusions}.


\section{Model setup}
\label{sec:model_setup}

To constrain the orientation of drifting grains under the effect of mechanical alignment, we applied 3D radiation hydrodynamical simulations \citep{Flock_etal_2017b, Flock_etal_2020} using the PLUTO code \citep{Mignone_etal_2007, Mignone_etal_2010, Mignone_etal_2012}.
Here, simulation snapshots provide the density of the gas and dust, as well as the velocity fields of the gas $\vec{\varv}_\mathrm{g}$ and the dust component $\vec{\varv}_\mathrm{d}$ in each cell.
We then define the drift velocity or headwind of the dust grains as
\begin{equation}
    \label{eq:drift_velocity}
    \vec{\varv}_\mathrm{drift} = \vec{\varv}_\mathrm{g} - \vec{\varv}_\mathrm{d}\, .
\end{equation}
In this study, we used the data of the simulation snapshot after 750 orbits of evolution.
After 650 orbits of evolution, particles of \SI{0.1}{mm} and\SI{1}{mm} in size were inserted at the midplane.
We let the particles evolve for a further 100 orbits to allow them to drift and mix into a new equilibrium state.
An overview of the setup parameter is described in Sect.~\ref{subsec:radiative_transfer} and the corresponding Table~\ref{tab:model_setup}.
For more information on the solved equations and the numerical setup, we refer to Sect.~2 in \citet{Flock_etal_2020}.

\subsection{Ballistic grain aggregation}
\label{subsec:grain_aggregation}

Modeling the mechanical alignment of dust is closely related to the shape and complex surface topology of individual grains.
Observations and modeling of (sub)millimeter polarization suggest the presence of compact grains in the disk midplane \citep{Stephens_etal_2020, Brunngraeber_Wolf_2021}, while scattering properties derived from mid-infrared observations suggest more elongated porous grains \citep{Ginski_etal_2023, Tanaka_etal_2023}.
The porous grain aggregates are likely to undergo some compression while settling toward the midplane \citep{Tanaka_etal_2023, Ueda_etal_2024}.

In this paper, we utilize moderately compact dust grains throughout the entire disk to estimate the parameters of mechanical alignment.
Dust grain aggregates are created using ballistic aggregation and migration (BAM) of primary particles (monomers).
Here, we follow the algorithm suggested in \citet{Shen_etal_2008}, with modifications presented in \citet{Reissl_etal_2024}, where monomers on random trajectories are successively added to the aggregate.
For this study, we allow for exactly one migration (BAM1) to establish additional connections, controlling the compactness of the aggregates.
The monomer radii are sampled from a log-normal distribution between \SI{10}{nm} and \SI{100}{nm} with a mean of \SI{20}{nm}.
The sampling from the log-normal distribution is biased to guarantee an exact pre-defined effective radius ${ a_{\mathrm{eff}} = \left( 3 \sum V_{\mathrm{mon}} / ( 4\pi ) \right)^{1/3} }$, where the total volume of the resulting aggregate is the sum of the volumes of the individual monomers $V_{\mathrm{mon}}$.

\begin{figure}
    \centering
    \includegraphics[width=0.91\linewidth]{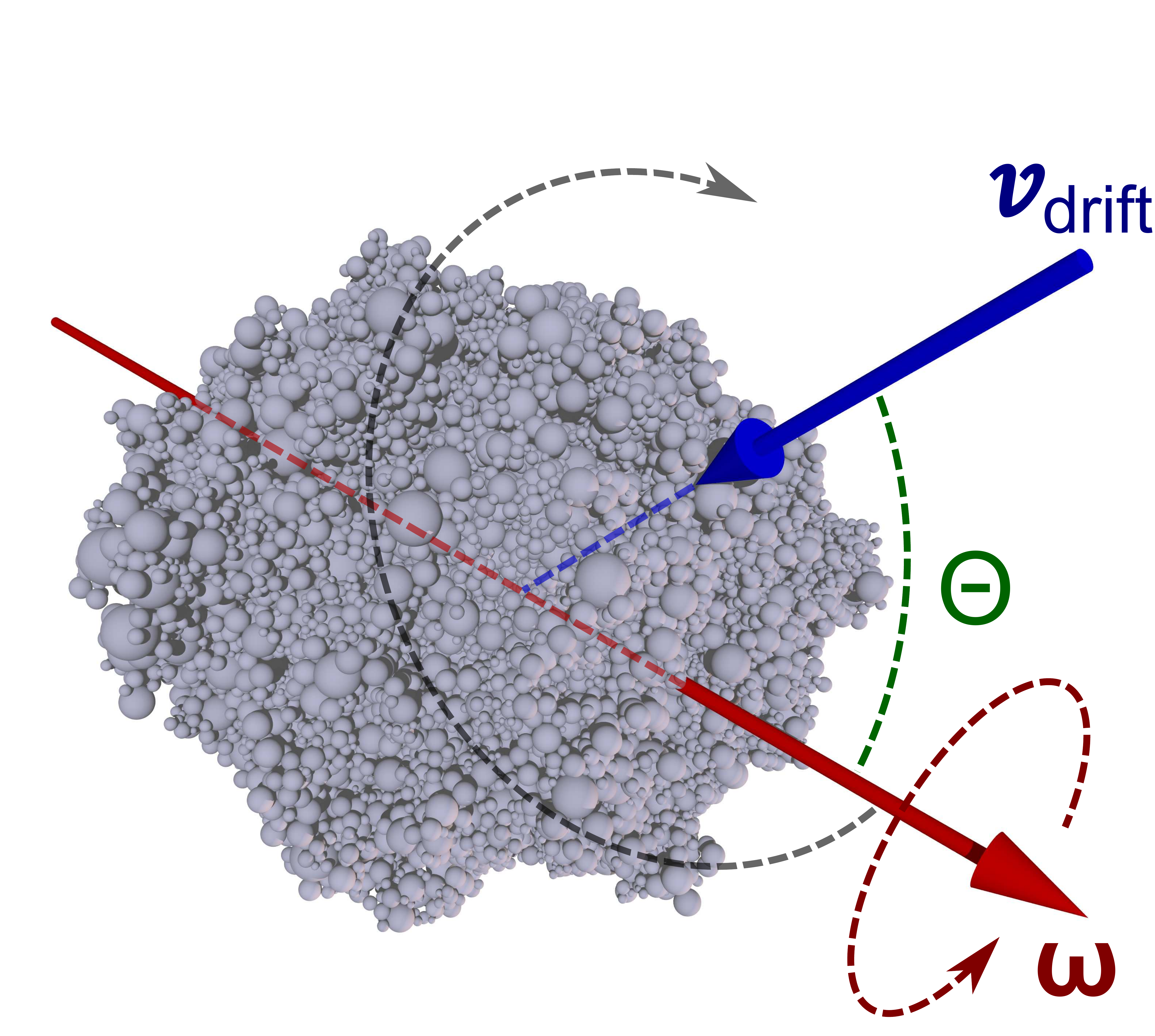}
    \caption{%
        Exemplary BAM1 dust aggregate with an effective radius of $a_{\mathrm{eff}} = \SI{1.2}{\um}$.
        The aggregate rotates with an angular velocity, $\vec{\omega,}$ that corresponds to its maximum moment of inertia, $I_{\mathrm{m}}$.
        The rotation and precession of the dust aggregate result from gas-dust interactions, where the gas and dust components move with a relative velocity, $\vec{\varv}_{\mathrm{drift}}$ , through the disk.
        The angle, $\Theta,$ is defined as the angle between $\vec{\omega}$ and $\vec{\varv}_{\mathrm{drift}}$.
    }
    \label{fig:MET_skecth}
\end{figure}

We pre-calculate aggregates with exact effective radii between \SI{0.1}{\um} and \SI{1.4}{\um}.
Calculating aggregates with larger effective radii $a_{\mathrm{eff}}$ is challenging due to increasing complexity within the BAM algorithm and resulting computational time constraints.
For each $a_{\mathrm{eff}}$, we use 30 random seeds to create a representative sample of BAM1 aggregates.
An exemplary BAM1 dust aggregate with $a_{\mathrm{eff}} = \SI{1.2}{\um}$ is depicted in Fig.~\ref{fig:MET_skecth}.

\subsection{Mechanical grain alignment}
\label{subsec:grain_alignment}

A gas-dust drift has long been suggested as a cause of mechanical alignment \citep{Gold_1952a, Gold_1952b}.
However, attempts to constrain the parameters of mechanical alignment lack predictive capability for a larger grain ensemble or do not sufficiently take the complex surface morphology of dust aggregates into account \citep[see, e.g.,][]{Lazarian_1994, Lazarian_Hoang_2007b, Das_Weingartner_2016, Hoang_etal_2018}.

In this paper, we apply the mechanical alignment of dust (MAD), a numerical framework for calculating alignment parameters using detailed Monte Carlo dust-gas interaction simulations.
The MAD framework considers the mechanical torques (MET) arising from scattering, sticking, and desorption of gas on the surface of fractal dust aggregates \citep[see][for further details]{Reissl_etal_2023}.
However, it is not a priory clear how the mechanical alignment of BAM1 aggregates would scale with an increasing grain size of $a_{\mathrm{eff}}$.
Hence, we explore the MAD for grain sizes of $a_{\mathrm{eff}} = \SI{0.1}{\um}$ to \SI{1.4}{\um} in order to derive a parametrized expression for the mechanical alignment.

In detail, the MAD is used to calculate the METs for each of the individual pre-calculated BAM1 aggregates introduced in Sect.~\ref{subsec:grain_aggregation} for representative values of the gas temperature, $T_{\mathrm{g}}$, gas number density, $n_{\mathrm{g}}$, dust temperature, $T_{\mathrm{d}}$, and the differential velocities of the gas component, $\vec{\varv}_\mathrm{g}$, and the dust component, $\vec{\varv}_\mathrm{d}$, separately.
The MAD simulation results are analyzed by utilizing the dimensionless quantity $\varv_\mathrm{drift} / \varv_\mathrm{th}$, that is, the absolute drift velocity, $\varv_\mathrm{drift} = \vert \vec{\varv}_\mathrm{drift} \vert$ normalized by the thermal velocity of the gas
\begin{equation}
    \label{eq:thermal_velocity}
    \varv_\mathrm{th} = \sqrt{\frac{2 k_\mathrm{B} T_\mathrm{g}}{\mu m_\mathrm{H}}}\, .
\end{equation}
Here, we assume that the gas temperature, $T_\mathrm{g}$, is equal to the average dust grain temperature and an average molecular weight of $\mu = 2$ assuming molecular hydrogen.

When modeling the rotational dynamics, the dust aggregate cannot be assumed to be a rigid body.
Internal dissipative processes \citep[see, e.g.,][]{Purcell_1979, Lazarian_Roberge_1997, Lazarian_Draine_1999, Lazarian_Efroimsky_1999} make, in the long run, the axis associated with the maximal moment of inertia, $I_{\mathrm{m}}$, the axis of rotation.
For simplicity, we assume the dissipation timescale of the dust aggregates to be small compared to the dynamical timescales of the disk.
Subsequently, the gas-dust interactions spin up a dust aggregate to an angular velocity $\vec{\omega}$.
We note that the METs acting on the dust aggregate would not align $\vec{\omega}$ parallel to the direction of the drift $\vec{\varv}_\mathrm{drift}$ but would rather cause a precession of $\vec{\omega}$ around $\vec{\varv}_\mathrm{drift}$ with an opening angle of $\Theta$.
A sketch of the alignment dynamics is provided in Fig.~\ref{fig:MET_skecth}.

For each set of input parameters ${ \{ n_{\mathrm{g}}, T_{\mathrm{g}}, T_{\mathrm{d}}, \varv_\mathrm{drift} / \varv_\mathrm{th} \} }$ we evaluate the dynamical evolution of the angular velocity $\vec{\omega}$ and the opening angle $\Theta$ via 
\begin{equation}
    \label{eq:time_evolution}
    \frac{\mathrm{d} \vec{\omega}}{\mathrm{d} t} =  \frac{1}{I_{\mathrm{m}}} \vec{\Gamma}_{\mathrm{MET}}\left(\Theta\right) - \frac{\vec{\omega} }{\tau_{\mathrm{drag}} }    
\end{equation}
for each BAM1 aggregate individually.
Here, $\vec{\Gamma}_{\mathrm{MET}}\left(\Theta\right)$ is the angle-dependent net MET, and $\tau_{\mathrm{drag}}$ is the characteristic timescale of the rotational drag, combining the effects of gas drag and IR drag due to photon emission \citep{Draine_Lazarian_1998}.
For the calculation of $\tau_{\mathrm{drag}}$ of BAM1 aggregates, we refer the reader to \citet{Reissl_etal_2023} and \citet{Jaeger_etal_2024}, respectively.
A solution to Eq.~\eqref{eq:time_evolution}, along with the average grain alignment behavior as a function of grain size and the corresponding timescales, is presented in Appendix~\ref{app:timescales}.

Furthermore, we trace the trajectory for each of the BAM1 aggregates with the same effective radius, $a_{\mathrm{eff}}$, in the $\omega-\Theta$ phase space defined by Eq.~\eqref{eq:time_evolution} to determine attractor points that are the long-term points of stability of the angular momentum, $\omega$, and the opening angle, $\Theta$.
As a lower limit for grain alignment, we take the thermal angular momentum ${ \omega_{\mathrm{th}} \approx \left(k_{\mathrm{B}} T_{\mathrm{d}} / I_{\mathrm{m}} \right)^{1/2} }$.
Aggregates rotating below $\omega < 3\, \omega_{\mathrm{th}}$ may still have attractors, but are regularly kicked out of stable alignment by random gas-dust collisions \citep{Hoang_Lazarian_2008} and, thus cannot contribute to the net dust polarization.

An upper limit arises from the fact that rapidly rotating dust grains may become disrupted by centrifugal forces, given a sufficiently high angular velocity $\omega_{\mathrm{disr}}$ \citep[see][for further details]{Hoang_etal_2019}.
Recently, a study by \citet{Reissl_etal_2024} based on N-body simulations of rotating aggregates provided a best-fit parameterization of $\omega_{\mathrm{disr}}$ based on basic material properties.
We utilize this parameterization to estimate the limit for rotational disruption, $\omega_{\mathrm{disr}}$, for each individual BAM1 grain, assuming typical material parameters of composite silicate grains as presented in \citet{Reissl_etal_2024}.
Using the exact grain composition introduced in Sect.~\ref{subsec:grain_aggregation} would require modeling dust aggregates where the water ice is not uniformly distributed within the aggregate but frozen to the aggregate's surface.
Such modeling with unevenly distributed materials goes beyond the scope of this paper.
However, given the variations in the N-body simulation results of \citet{Reissl_etal_2024} for different grain compositions, we estimate that the magnitude of $\omega_{\mathrm{disr}}$ may be off by a factor of two at most.

Finally, we introduce the ratio, $\zeta$.
This ratio quantifies the number of BAM1 aggregates that have an attractor point within ${ 3\, \omega_{\mathrm{th}} < \omega < \omega_{\mathrm{disr}} }$ relative to the total number of pre-calculated aggregates per effective radius, $a_{\mathrm{eff}}$, to quantify the dust polarization of an ensemble of dust grains \citep[see][for further details]{Reissl_etal_2023}.
The maximum possible dust polarization is determined by the refractive index and the shape of the dust grain.
However, due to the precession of the dust grains, the maximal polarization becomes further reduced.
This reduction is usually quantified by the Rayleigh reduction factor \citep[see, e.g.,][]{Greenberg_1968, Lee_Draine_1985, Roberge_Lazarian_1999}
\begin{equation}
    \mathcal{R} = \frac{3\zeta}{2}\left( \langle \cos^2 \Theta \rangle - \frac{1}{3} \right)\, ,
    \label{eq:RRF}
\end{equation}
where $\langle \cos^2 \Theta \rangle$ is the ensemble average of all the angles $\Theta$ of the individual BAM1 aggregates at their attractor points.
We emphasize that we use a definition of the Rayleigh reduction factor modified by the ratio $\zeta$.
Exemplary distributions of $\Theta$ and the corresponding values of $\mathcal{R}$ are presented in Appendix~\ref{app:alignment} for selected grain ensembles of similar size.

\begin{figure*}
    \centering
    \includegraphics[width=\linewidth]{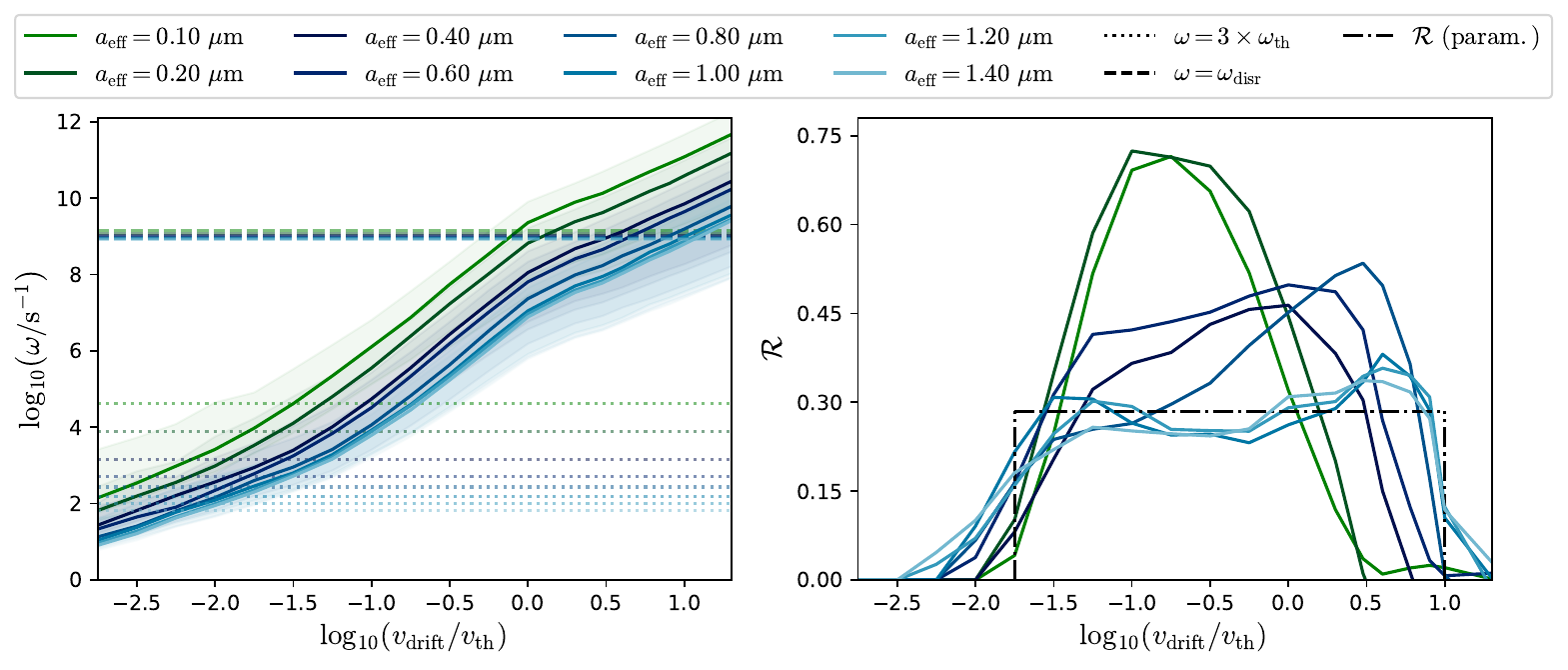}
    \caption{%
        Left panel: Angular velocity, $\omega,$ as a function of the ratio ${\varv_{\mathrm{drift}} / \varv_{\mathrm{th}}}$.
        The effective grain radii, $a_{\mathrm{eff}} = \SI{0.1}{\um}$ to \SI{1.4}{\um}, are color-coded.
        Solid lines represent the average values of $\omega$, while the shaded areas indicate the range between the minimum and maximum values.
        Dotted lines show the lower limit, $3\times \omega_{\mathrm{th}}$, for long-term stable alignment, and dashed lines represent the upper limit for rotational disruption, $\omega_{\mathrm{disr}}$.
        Right panel: Same as the left panel, but for the Rayleigh reduction factor $\mathcal{R}$ (solid lines).
        Dashed-dotted lines depict the parametric representation of $\mathcal{R}$ for grains with $a_{\mathrm{eff}} \geq \SI{1}{\um}$.
        See Sect.~\ref{subsec:grain_alignment} for details.
    }
    \label{fig:MET_alignment}
\end{figure*}

What the MAD Monte Carlo simulations in \citet{Reissl_etal_2023} reveal is that, by increasing the gas density, $n_{\mathrm{g}}$, the MET, $\Gamma_{\mathrm{MET}}$, increases, as does the gas drag.
Beyond a certain critical density, $n_{\mathrm{c}} \approx \SI{e2}{cm^{-3}}$, the MET and gas drag reach a balance, and the angular velocity $\omega$ of the aggregates reaches its terminal value.
For the range of gas densities $n_{\mathrm{g}} \approx \SI{e7}{cm^{-3}}$ to \SI{e11}{cm^{-3}} present in our simulation data, each individual dust grain is already beyond this critical limit $n_{\mathrm{c}}$.
Therefore, the grain alignment process is considered to be independent of $n_{\mathrm{g}}$ for this particular PLUTO simulation.

The same applies to the gas temperature, $T_{\mathrm{g}}$, and the dust temperature, $T_{\mathrm{d}}$.
An increase in $T_{\mathrm{g}}$ raises the impact velocity (see Eq.~\ref{eq:thermal_velocity}) and subsequently increases $\Gamma_{\mathrm{MET}}$.
With an increase in $T_{\mathrm{d}}$, IR photons carry away a fraction of the rotational energy, slowing down the grain rotation.
However, since both temperatures, $T_{\mathrm{g}}$ and $T_{\mathrm{d}}$, are below \SI{100}{K} (compare  Fig.~\ref{fig:avg_dust_temp}), their impact on the overall grain rotation dynamics is minimal, making the ratio $\varv_{\mathrm{drift}} / \varv_{\mathrm{th}}$ the dominant parameter for modeling dust rotation and subsequent polarization.

In Fig.~\ref{fig:MET_alignment}, we present the average angular momentum $\omega$ as a function of the ratio $\varv_{\mathrm{drift}} / \varv_{\mathrm{th}}$ under typical disk conditions:
$n_{\mathrm{g}} = \SI{e8}{cm^{-3}}$, $T_{\mathrm{g}} = \SI{20}{K}$, and $T_{\mathrm{d}} = \SI{20}{K}$, with effective grain radii $a_{\mathrm{eff}} = \SI{0.1}{\um}$ to \SI{1.4}{\um}.
As drift velocity, $\varv_{\mathrm{drift}}$, increases, the aggregates experience collisions predominantly from the direction of $\vec{\varv}_{\mathrm{drift}}$ rather than from random directions.
This results in a systematic mechanical torque $\Gamma_{\mathrm{MET}}$ with a steady orientation over time.
Additionally, an increase in $\varv_{\mathrm{drift}}$ raises $\Gamma_{\mathrm{MET}}$ due to a higher gas-dust collision rate.
The resulting average angular momentum $\omega$ for aggregates with $a_{\mathrm{eff}} = \SI{0.1}{\um}$ surpasses the threshold for stable alignment, $3 \, \omega_{\mathrm{th}}$, at $\varv_{\mathrm{drift}} \approx 0.03\, \varv_{\mathrm{th}}$, and such aggregates become rotationally disrupted on average at $\varv_{\mathrm{drift}}\approx \varv_{\mathrm{th}}$.

We emphasize that individual grains may still have stable alignment attractors within the range ${ 3 \, \omega_{\mathrm{th}} < \omega < \omega_{\mathrm{disr}} }$ even though the average angular velocity, $\omega$, does not meet this criterion.
Thus, the Rayleigh reduction factor $\mathcal{R} \approx 0$ to 0.75 for the ensemble of the smallest grains within the range ${ 0.01 \lesssim \varv_{\mathrm{drift}} / \varv_{\mathrm{th}} \lesssim 3 }$.
For the largest aggregates, with $a_{\mathrm{eff}} = \SI{1.4}{\um}$, the thermal angular momentum $\omega_{\mathrm{th}}$ is lower due to their larger moment of inertia, $I_{\mathrm{m}}$, while the limit $\omega_{\mathrm{th}}$ remains in the same order of magnitude for all aggregates \citep[see][]{Reissl_etal_2024}.

As reported in \citet{Reissl_etal_2023}, larger grains are more efficiently spun up due to their increased surface area, whereas for grains with rounder shapes, the net mechanical torque cancels out over time.
For the specific BAM1 aggregates used in this study, smaller grains are more elongated, while the shapes of larger grains become increasingly spherical \citep{Reissl_etal_2024}.
Considering both factors -- namely, the increase in surface area countered by the reduced net torque due to a rounder shape -- the average behavior of the BAM1 aggregates becomes more uniform with increasing radius $a_{\mathrm{eff}}$.
What we report here is that the mechanical alignment behavior converges for grain sizes $a_{\mathrm{eff}} \geq \SI{1.0}{\um}$.
Consequently, our approach to model the mechanical alignment for individual grain sizes of $a_{\mathrm{eff}} = \SI{0.1}{\um}$ to \SI{1.4}{\um} allows us to extrapolate the alignment behavior up to a grain size of \SI{100}{\um} and \SI{1}{mm}, respectively (see Table~\ref{tab:model_setup}).
As shown in Fig.~\ref{fig:MET_alignment}, this results in a Rayleigh reduction factor that can be approximated by
\begin{equation}
    \label{eq:rayleigh_reduction_factor}
    \mathcal{R} =
        \begin{cases}
            0.29 & \text{if}\ \num{1.78e-2} \leq \varv_\mathrm{drift} / \varv_\mathrm{th} \leq 10\, , \\
            0 & \text{otherwise}
        \end{cases}
\end{equation}
for larger grains.
For simplicity, we used this approximation of $\mathcal{R}$ for all dust grains with $a_{\mathrm{eff}} \geq \SI{1}{\um}$ in the subsequent RT simulations.
Here, the velocity fields of the gas and each dust grain in the protoplanetary disk are provided by the simulation snapshot (see Appendix~\ref{app:velocity_field}).
Consequently, the direction of alignment is determined by the direction of the drift velocity $\vec{\varv}_{\mathrm{drift}}$, where helical grains align with their short axis parallel to the drift velocity vector.

\subsection{Optical properties of dust grains}
\label{subsec:grain_properties}

To calculate the optical properties of dust grains of different sizes at various wavelengths in a reasonable amount of time, we ignore any porosity or irregularly shaped particles.
In particular, calculation of the optical properties for various incident and scattering angles for the exemplary dust aggregate shown in Fig.~\ref{fig:MET_skecth} would be computationally too time-consuming and is out of the scope of the current study.
We refer the reader to \citet{Kirchschlager_Wolf_2014} and \citet{Kirchschlager_etal_2019} who discussed the impact of porosity on the polarization degree, or to \citet{Kirchschlager_Bertrang_2020} who discussed the scattering of nonspherical dust grains on the polarization.
Instead, we assume compact spherical grains for dust grain heating and polarization as a result of scattering and elongated oblate-shaped grains for polarization as a result of dichroic emission and absorption.

\begin{figure}
    \centering
    \includegraphics[width=\linewidth]{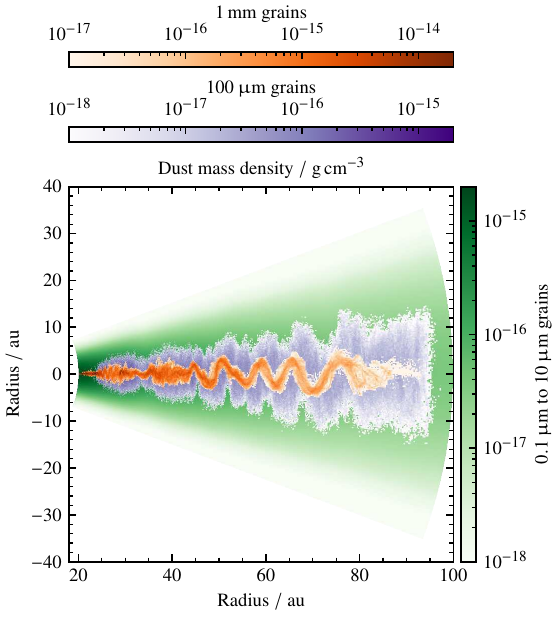}
    \caption{%
        Cross-section view of the dust mass density in units of \si{g.cm^{-3}} of \SI{0.1}{\um} to \SI{10}{\um}-sized grains (green), \SI{100}{\um}-sized grains (purple), and \SI{1}{mm}-sized grains (orange).
        The values are azimuthally averaged.
    }
    \label{fig:dust_mass_dens}
\end{figure}

The dust grains are divided into three size bins.
The first bin represents small grains with sizes of \SI{0.1}{\um} to \SI{10}{\um} distributed with a power law,
\begin{equation}
    N(a) \propto a^{-3.5}\, .
\end{equation}
These dust grains are fully coupled to the dynamics of the gas during hydrodynamical simulations and thus are not affected by mechanical grain alignment in this study.
The other two bins represent larger grains with a log-normal grain size distribution,
\begin{equation}
    N(a) \propto \frac{1}{a} \exp\left( -\frac{(\ln(a / a_\mathrm{eff}))^2}{2\sigma^2} \right)\, ,
\end{equation}
where $\sigma = 0.25$ and $a_\mathrm{eff}$ is either \SI{100}{\um} or \SI{1}{mm}.
These grains are less coupled to the gas movement, thus, experience a headwind from the gas and are able to align in the protoplanetary disk.
We note that in the 3D hydrodynamical simulations, only single grain sizes of \SI{100}{\um} and \SI{1}{mm} are applied.
For polarized radiative transfer, we take a size distribution since dust grains in circumstellar disks occur in different sizes and, thus, contributions of single grains are smoothed or cancel out.
In Fig.~\ref{fig:dust_mass_dens}, we present the dust mass density distribution of the different grain sizes resulting from a simulation using the PLUTO code.
The wave pattern of the large dust grains is due to the large-scale upward and downward gas movement by the VSI \citep{Flores-Rivera_etal_2020}.

We apply the DSHARP dust composition \citep{Birnstiel_etal_2018}, consisting of refractory organics \citep{Henning_Stognienko_1996} with a fraction of \SI{40}{\percent} by mass, \SI{33}{\percent} astronomical silicate \citep{Draine_2003}, \SI{20}{\percent} water ice \citep{Warren_Brandt_2008}, and \SI{7}{\percent} troilite \citep{Henning_Stognienko_1996}.
The composition is mixed using the formula of \citet{Bruggeman_1935}, and the resulting material density is \SI{1.675}{g.cm^{-3}}.

\begin{figure}
    \centering
    \includegraphics[width=\linewidth]{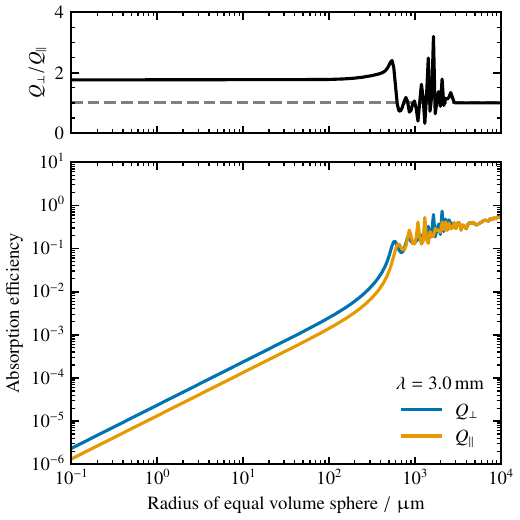}
    \caption{%
        Ratio of absorption efficiencies $Q_\perp / Q_\parallel$ (top) and absorption efficiency (bottom) of an oblate spheroid as a function of radius of an equal volume sphere.
        This figure shows an axis ratio of the spheroid of $1 / 1.5$ and a wavelength of radiation of \SI{3}{mm}.
        The absorption efficiency $Q_\parallel$ is the efficiency along the minor grain axis, that is, the axis of alignment.
        If $a / \lambda \geq 1$, we assume a spherical particle, thus, $Q_\perp = Q_\parallel$.
        See Sect.~\ref{subsec:grain_properties} for details.
    }
    \label{fig:absorption_eff}
\end{figure}

For spherical particles, the optical properties of dust grains are calculated using the Mie scattering theory \citep{Mie_1908} and the code MIEX \citep{Wolf_Voshchinnikov_2004, Wolf_Voshchinnikov_2018}.
Optical properties for nonspherical \SI{100}{\um} and \SI{1}{mm}-sized dust grains are calculated using the discrete dipole approximation \citep{Draine_Flatau_1994} and the code DDSCAT 7.3.3 \citep{Draine_Flatau_2000, Draine_Flatau_2008}.\footnote{%
    For calculations of cross sections, we used \num{281250} dipoles, which corresponds to an upper limit of $a / \lambda \lesssim 2.8$ \citep{Draine_Flatau_2013}.
}
Here, we assume that the spheroidal particles are oblate with a ratio of the minor to major axes of $1 / 1.5$.
For large dust grains, that is, $a \gtrsim \lambda / 2$, where $\lambda$ is the wavelength of radiation, the polarized emission becomes unpolarized even if the grains are aligned \citep{Cho_Lazarian_2007, Kirchschlager_etal_2019}.
Therefore, if $a / \lambda \geq 1$, instead of using DDSCAT, we assume spherical particles and calculate the optical properties with MIEX.
For the largest \SI{1}{mm}-sized grains, we obtain an average absorption mass opacity of approximately \SI{0.78}{cm^2.g^{-1}} and an average scattering mass opacity of approximately \SI{11.9}{cm^2.g^{-1}} at a wavelength of \SI{3}{mm}.

Figure~\ref{fig:absorption_eff} shows the absorption efficiency of an oblate spheroid as a function of radius of an equal volume sphere at a wavelength of \SI{3}{mm}.
If the grain size is small compared to the wavelength, that is, if $a / \lambda \lesssim 0.1$, the ratio of the absorption efficiency is approximately 1.8.
With increasing grain size, that is, $2 \pi a \gtrsim \lambda$, the cross sections of single grain sizes show oscillations due to multiple internal reflections of radiation \citep{van_de_Hulst_1981}.
This also causes a ratio of absorption efficiency below 1 and a change in the orientation of polarization \citep[see][]{Cho_Lazarian_2007, Kirchschlager_etal_2019, Guillet_etal_2020}.
Moreover, porosity of dust grains would reduce the intensity of oscillations in the cross sections in the Mie regime ($2 \pi a \sim \lambda$) compared to compact dust particles.

\subsection{Radiative transfer}
\label{subsec:radiative_transfer}

The prediction and analysis of multi-wavelength polarization measurements requires a comprehensive understanding of radiation transfer, dust properties, and underlying physical mechanisms such as grain alignment.
To model synthetic observations of protoplanetary disks, we apply the 3D Monte Carlo radiative transfer code POLARIS\footnote{\url{https://github.com/polaris-MCRT/POLARIS}} \citep{Reissl_etal_2016, Reissl_etal_2018}.
The model parameters are summarized in Table~\ref{tab:model_setup}.
The stellar parameters are adopted from \citet{Flock_etal_2020} and \citet{Blanco_etal_2021}.

\begin{table}

    \centering
    \caption{%
        Model parameters used for the numerical simulations.
    }
    \label{tab:model_setup}
    \renewcommand{\arraystretch}{1.2}
    \small
    \begin{tabular*}{\linewidth}{@{\extracolsep{\fill}} l r}
        \hline\hline
        Parameter & Value \\
        \hline
        Stellar radius & \SI{2}{R_\odot} \\
        Stellar mass & \SI{0.5}{M_\odot} \\
        Effective temperature & \SI{4000}{K} \\
        Stellar luminosity & \SI{0.92}{L_\odot} \\
        Inner disk radius & \SI{20}{au} \\
        Outer disk radius & \SI{100}{au} \\
        PLUTO grid resolution ($r, \theta, \phi$) & $1024 \times 512 \times 2044$ \\
        Total gas mass & \SI{2.4e-02}{M_\odot} \\
        Dust mass (small grains) & \SI{8.0e-05}{M_\odot} \\
        Dust mass (\SI{100}{\um} grains) & \SI{1.7e-05}{M_\odot} \\
        Dust mass (\SI{1}{mm} grains) & \SI{5.4e-05}{M_\odot}\\
        Distance & \SI{140}{pc} \\
        \hline
    \end{tabular*}

\end{table}

First, based on the dust density distribution, POLARIS calculates the temperature of the dust grains that are heated by the central stellar radiation source.
The resulting temperature distribution is shown in the Appendix~\ref{app:temperature_distribution}.
Subsequently, the polarized radiation from dichroic emission and absorption for aligned oblate spheroids was calculated.
However, the dust grains are not perfectly aligned, causing a decrease in the observed net polarization compared to that expected for the individual grains.
To account for this effect, we use the Rayleigh reduction factor defined in Eq.~\eqref{eq:rayleigh_reduction_factor}.
In this work, we consider either \SI{100}{\um} or \SI{1}{mm} grains aligned, and we study each of their contributions to the polarization at different wavelengths.
The respective other grain size, as well as the smaller grains (\SI{0.1}{\um} to \SI{10}{\um}) have a random orientation.
Finally, the polarized scattered radiation of the star and dust, that is, self-scattering, is computed by assuming spherical dust grains.
In this step, the contribution of all grains to the net polarization is taken into account.

To describe the state and degree of polarization of the radiation, we use the Stokes formalism \citep{Bohren_Huffman_1998}.
Here, the radiation is characterized by its total flux $I$, the components $Q$ and $U$, which describe the linear polarization, and the component $V$ for the circular polarization.
The degree $P$ and orientation $\psi$ of linear polarization are defined by
\begin{equation}
    P = \frac{\sqrt{Q^2 + U^2}}{I}\, , \quad
    \tan(2\psi) = \frac{U}{Q}\, .
\end{equation}
For a detailed description of the computational methods, we refer the reader to \citet{Reissl_etal_2016} and Appendix B in \citet{Reissl_etal_2019}.


\section{Results}
\label{sec:results}

\begin{figure}
    \centering
    \includegraphics[width=\linewidth]{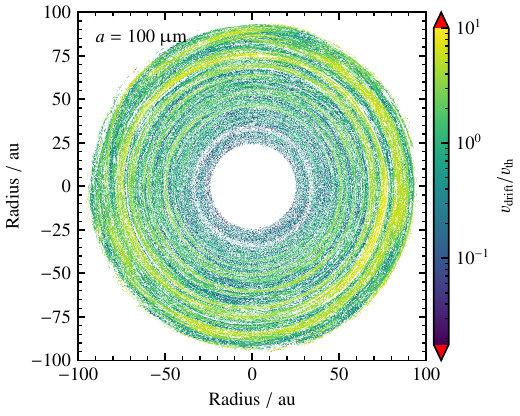}
    \includegraphics[width=\linewidth]{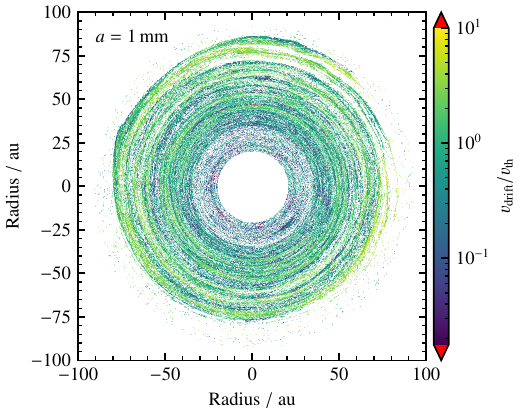}
    \caption{%
        Ratio of the drift velocity and the thermal velocity in the disk midplane of \SI{100}{\um} (top) and \SI{1}{mm}-sized (bottom) grains.
        Dust grains are aligned if $\num{1.78e-2} \leq \varv_\mathrm{drift} / \varv_\mathrm{th} \leq 10$ (see Eq.~\ref{eq:rayleigh_reduction_factor}), otherwise they are marked in red in the figure.
        See Sect.~\ref{subsec:orientation_alignment} for details.
    }
    \label{fig:thermal_vel_xy}
\end{figure}

Before calculating the flux and polarization maps, we have to determine whether grains are aligned in the protoplanetary disk and, in that case, the orientation of their axis of alignment.
Most importantly, this is based on the drift velocity and temperature at each location in the protoplanetary disk.
Subsequently, we analyze the contributions of the thermally reemitted radiation and dichroic extinction of the aligned grains as well as of self-scattered radiation to the polarization.

\subsection{Orientation of aligned grains}
\label{subsec:orientation_alignment}

\begin{figure*}
    \centering
    \begin{minipage}{0.49\linewidth}
        \centering
        \includegraphics[width=\linewidth]{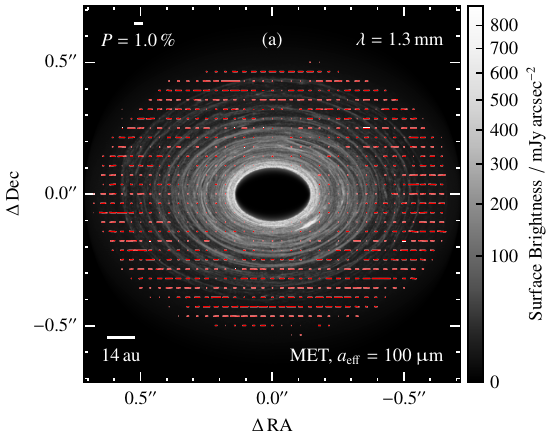}
    \end{minipage}
    \begin{minipage}{0.49\linewidth}
        \centering
        \includegraphics[width=\linewidth]{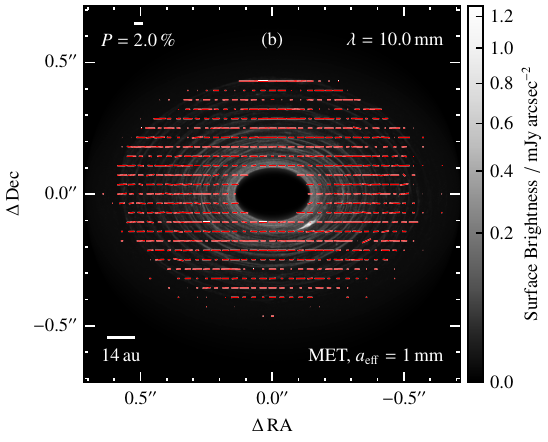}
    \end{minipage}
    \begin{minipage}{0.49\linewidth}
        \centering
        \includegraphics[width=\linewidth]{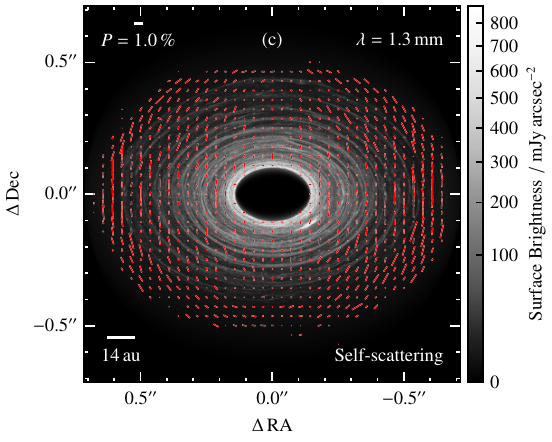}
    \end{minipage}
    \begin{minipage}{0.49\linewidth}
        \centering
        \includegraphics[width=\linewidth]{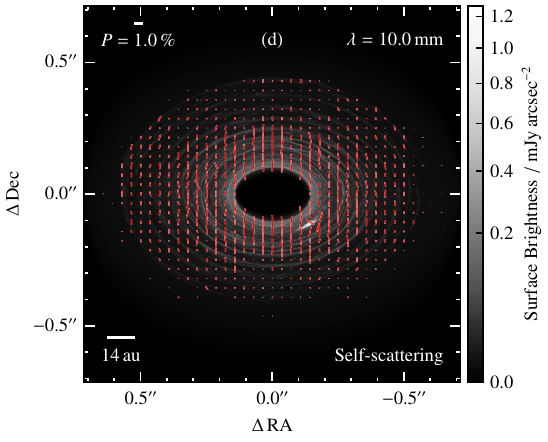}
    \end{minipage}
    \caption{%
        Surface brightness map of the total emitted thermal radiation at a wavelength of \SI{1.3}{mm} (left) and \SI{10}{mm} (right) overlaid with polarization vectors at these wavelengths.
        In the top row (a, b), the polarization arises from thermal emission and dichroic extinction of aligned dust grains only.
        At wavelengths of \SI{1.3}{mm} and \SI{10}{mm}, only \SI{100}{\um} and \SI{1}{mm}-sized grains are assumed to be aligned, respectively.
        In contrast, in the bottom row (c, d), the polarization arises from self-scattered radiation only.
        The disk is inclined by \ang{45}.
        See Sects.~\ref{subsec:polarization_alignment} and~\ref{subsec:polarization_scattering} for details.
    }
    \label{fig:sb_align_selfsca_i45}
\end{figure*}

Figure~\ref{fig:thermal_vel_xy} shows the ratio of the drift velocity and the thermal velocity (see Eq.~\ref{eq:rayleigh_reduction_factor}) of \SI{100}{\um} and \SI{1}{mm}-sized grains in the disk midplane.
Regardless of the grain size and distance to the center, most dust grains meet Eq.~\eqref{eq:rayleigh_reduction_factor} and are therefore aligned due to METs.
The minor axis of the oblate grain is then oriented parallel to the drift velocity vector $\vec{\varv}_\mathrm{drift}$.
Closer to the inner rim of the disk, there is a pattern of changing velocity ratio $\varv_\mathrm{drift} / \varv_\mathrm{th}$, in particular for grains with a size of \SI{1}{mm}.
This can be traced back to an alternating velocity field close to the inner radius of the disk (see Appendix~\ref{app:velocity_field}) and is a numerical effect, as the region inward \SI{25}{au} is affected by the boundary conditions.
In contrast, an interesting feature is the change of alignment of grains trapped in a vortex.
There, the velocity ratio decreases to the lower limit of Eq.~\eqref{eq:rayleigh_reduction_factor}, in contrast to the local environment.

Due to the VSI, the dominant component of the drift velocity is parallel to the rotational disk axis (see Appendix~\ref{app:velocity_field}).
Thus, most of the elongated grains are oriented with their short axis perpendicular to the disk midplane, since we assume that the grains align with their short axis parallel to the drift velocity vector.
As a consequence, the polarization is close to zero for a face-on disk since, in this case, the main axis of grain alignment is parallel to the line of sight, and oblate-shaped particles appear symmetric for this viewing geometry.
Therefore, we consider an inclined disk for the following analysis.

\subsection{Polarization of aligned grains}
\label{subsec:polarization_alignment}

Figures~\ref{fig:sb_align_selfsca_i45}a and~\ref{fig:sb_align_selfsca_i45}b show the surface brightness map of the thermal reemission at wavelengths of \SI{1.3}{mm} and \SI{10}{mm}, respectively.
The disk has an inclination of \ang{45}.
We assume that at wavelengths of \SI{1.3}{mm} and \SI{10}{mm}, only grains with a size of \SI{100}{\um} and \SI{1}{mm} are aligned due to METs and contribute to the polarization, respectively.

In general, the thermal reemission flux decreases with increasing distance to the center and with increasing wavelength.
Due to the radially inhomogeneous distribution of dust grains, there are gaps and rings visible on the surface brightness maps.
Furthermore, there is a local increase in surface brightness caused by the accumulation of dust grains due to a vortex formed at a position angle of approximately \ang{225}.
At a wavelength of \SI{10}{mm}, the contrast in surface brightness with the surroundings increases even more and is a prominent feature there.
We also refer the reader to \citet{Blanco_etal_2021} who discussed the impact of substructures produced by the VSI on synthetic photometric observations.

Figure~\ref{fig:sb_align_selfsca_i45} also shows the degree and orientation of polarization.
Hereby, the direction of polarization is parallel to the major axis of the disk.
This can be explained by the drift velocity vector, which has its largest component in the polar direction, that is, perpendicular to the disk midplane (see Sect.~\ref{subsec:orientation_alignment}).

\begin{figure}
    \centering
    \includegraphics[width=\linewidth]{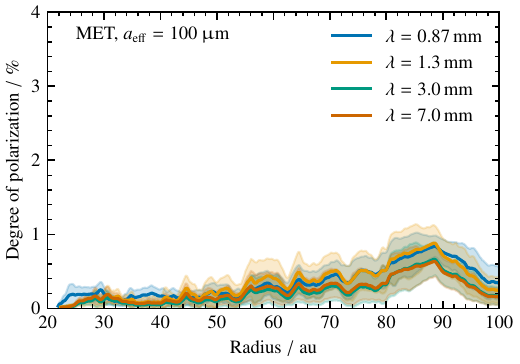}
    \includegraphics[width=\linewidth]{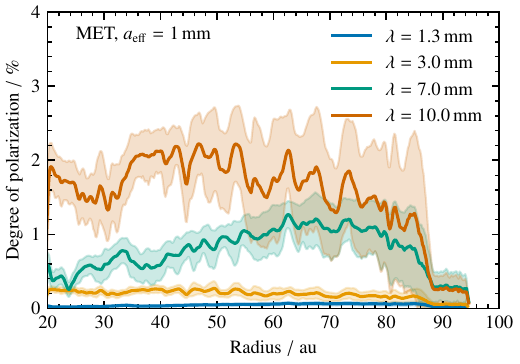}
    \includegraphics[width=\linewidth]{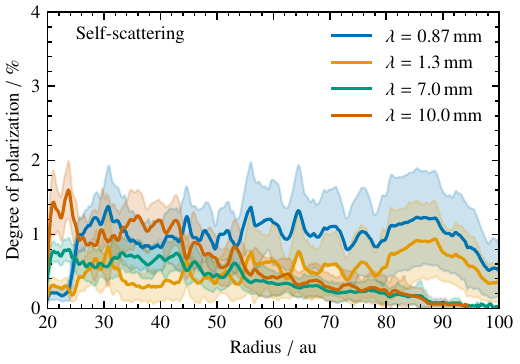}
    \caption{%
        Wavelength dependence of the radial profile of the polarization degree.
        In the top and middle figure, only \SI{100}{\um} and \SI{1}{mm}-sized grains are aligned and contribute to the polarization, respectively.
        In the bottom figure, the polarization arises from self-scattered radiation only.
        The disk is inclined by \ang{45}.
        See Sects.~\ref{subsec:polarization_alignment} and~\ref{subsec:polarization_scattering} for details.
    }
    \label{fig:pol_rad_i45}
\end{figure}

In the next step, we investigate the radial dependence of the degree of polarization.
For this purpose, the polarization map of the inclined disk was deprojected and then averaged azimuthally to obtain the radial profile (see Fig.~\ref{fig:pol_rad_i45}).
This is justified since the polarization profile along the major and minor axes is identical for our model setup and the considered wavelengths.

If only \SI{100}{\um}-sized grains are aligned, the degree of polarization is small throughout the disk, which are values below \SI{1}{\percent}.
This is due to the decreasing absorption efficiency with decreasing dust grain size (see Fig.~\ref{fig:absorption_eff}), thus, the thermal emission and polarized radiation is dominated by the larger \SI{1}{mm}-sized grains.
In addition, the degree of polarization increases with increasing distance to the center.
This finding can be explained by the large millimeter-sized grains that are settled to the midplane and migrated inward to the inner rim of the disk (see Fig.~\ref{fig:dust_mass_dens}).

If only \SI{1}{mm}-sized grains are aligned, the degree of polarization significantly depends on the considered wavelength.
In particular, for a wavelength of \SI{10}{mm}, the degree of polarization increases to about \SI{2.5}{\percent}.
In contrast, for a wavelength of \SI{1.3}{mm}, the degree of polarization is below \SI{0.2}{\percent} throughout the disk.
Moreover, the radial polarization profile decreases with increasing radius for wavelengths of \SI{3}{mm} and \SI{10}{mm}.
This can also be explained by the large millimeter-sized grains that settled to the midplane and migrated inward to the inner rim of the disk.
As a consequence, the polarization due to thermally emitted radiation decreases with increasing radius.

In addition, there are multiple maxima and minima in the polarization profiles.
These are caused by gaps and rings in the surface brightness, and thus minima and maxima in the dust density distribution.
Depending on the total optical depth of the structure and thus the observing wavelength, a gap or bright rings cause either an increase or decrease in the total degree of linear polarization.
Most interestingly, at a wavelength of \SI{7}{mm}, the degree of polarization also decreases closer to the inner rim of the disk, which is in contrast to the outlined behavior at wavelengths of \SI{3}{mm} and \SI{10}{mm}.
Finally, at a wavelength of \SI{10}{mm} and a radius of about \SI{30}{au}, the degree of polarization has a local minimum.
This is the result of the formed vortex, which has a lower degree of polarization compared to regions with the same distance to the center.

\subsection{Polarization due to self-scattering}
\label{subsec:polarization_scattering}

Polarization arises not only from emission by aligned grains, but also from radiation emitted and scattered by dust grains before leaving the disk (self-scattering).
Figures~\ref{fig:sb_align_selfsca_i45}c and~\ref{fig:sb_align_selfsca_i45}d shows the surface brightness map of the thermal radiation at a wavelength of \SI{1.3}{mm} and \SI{10}{mm}, respectively.
Again, the disk has an inclination of \ang{45}.
Polarization is caused only by self-scattered radiation, whereas grains of all sizes contribute to the net polarization.

In general, scattered radiation contributes significantly to the observed polarization at these wavelengths, resulting in a degree of polarization of about $\lesssim$\SI{1.5}{\percent} for all considered wavelengths.
Here, the highest degree of polarization is obtained for the shortest considered wavelength of \SI{0.87}{mm}.
With increasing wavelength, thus decreasing optical depth, the contribution of self-scattering decreases as well.
However, since most \SI{1}{mm}-sized particles have already migrated inward, polarization due to self-scattering has the greatest contribution at the inner disk rim, in particular at a wavelength of \SI{10}{mm} (see Fig.~\ref{fig:sb_align_selfsca_i45}).

The orientation of the resulting polarization is usually parallel to the projected minor disk axis.
This is caused by the anisotropic radiation field and the fact that scattered radiation is polarized perpendicular to the scattering plane, in particular, for grains that are smaller than the wavelength (Rayleigh limit).
On the other hand, for a face-on disk, the polarization vectors would preferentially be averaged azimuthally.
We refer to \citet{Kataoka_etal_2015} and \citet{Yang_etal_2016} who discuss this mechanism for thermal radiation in detail.
Note that the effect of polarization reversal (see Sect.~\ref{subsec:flip_polarization}), as discussed by \citet{Yang_etal_2016} and \citet{Brunngraeber_Wolf_2019}, may cause a radial polarization pattern.

Figure~\ref{fig:pol_rad_i45} also shows the wavelength dependence of the radial profile of the degree of polarization due to self-scattering.
Again, to derive this profile, the polarization map of the inclined disk was deprojected and then azimuthally averaged.
We note that in contrast to the polarization of thermal emission, the polarization due to self-scattering slightly differs between the projected major and minor axes of the protoplanetary disk.
In particular, the degree of self-scattering polarization is larger along the major axis of the disk compared to the minor axis of the disk \citep[see][]{Yang_etal_2016}.

Most importantly, in Fig.~\ref{fig:pol_rad_i45}, the degree of polarization of self-scattered is in the range or even exceeds the polarization due to emission of aligned \SI{100}{\um}-sized grains.
As a consequence, polarization due to self-scattering is the dominant polarization mechanism in this case.
In contrast, polarized emission of aligned grains with a size of \SI{1}{mm}, in particular at wavelengths of \SI{7}{mm} and \SI{10}{mm}, exceeds the polarization due to self-scattered radiation.
Furthermore, the polarization profile decreases with increasing radius at wavelengths of \SI{7}{mm} and \SI{10}{mm}.
Similarly to polarized emission, this is caused by large millimeter-sized grains that settled down to the midplane and migrated inward to the inner rim of the disk (see Fig.~\ref{fig:dust_mass_dens}).

\subsection{Polarization near the vortex}
\label{subsec:polarization_vortex}

\begin{figure}
    \centering
    \includegraphics[width=\linewidth]{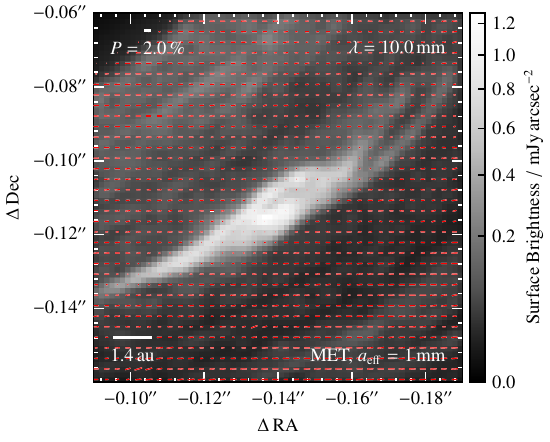}
    \includegraphics[width=\linewidth]{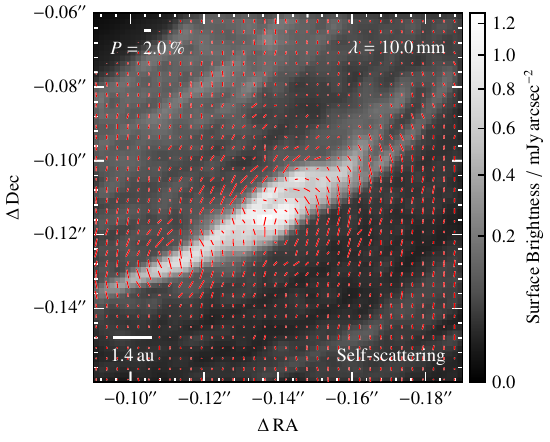}
    \caption{%
        Surface brightness map at a wavelength of \SI{10}{mm} with an enlarged image section around the vortex.
        The polarization arises from thermal emission and dichroic absorption of aligned \SI{1}{mm}-sized dust grains (top) and from self-scattered radiation (bottom) only.
        The disk is inclined by \ang{45}.
        See Sect.~\ref{subsec:polarization_vortex} for details.
    }
    \label{fig:sb_a1mm_selfsca_i45_w10mm_zoom}
\end{figure}

In the vicinity of the above-mentioned vortex, not only the surface brightness increases but the polarization degree changes as well.
In particular, the polarization due to the emission of thermal radiation of aligned dust grains decreases with increasing surface brightness.
This is shown in Fig.~\ref{fig:sb_a1mm_selfsca_i45_w10mm_zoom}, which enlarges the section around the formed vortex.
In contrast to the local environment, the dust density and thus the optical depth increase there.
Consequently, the polarization due to the emission of thermal radiation and extinction of aligned dust grains cancel each other out.
For the case of self-scattering, the resulting polarization vectors do not show a homogeneous pattern anymore due to the inhomogeneous radiation field.
Moreover, the degree of polarization increases with increasing surface brightness.

\subsection{Flip of the polarization orientation}
\label{subsec:flip_polarization}

\begin{figure}
    \centering
    \includegraphics[width=\linewidth]{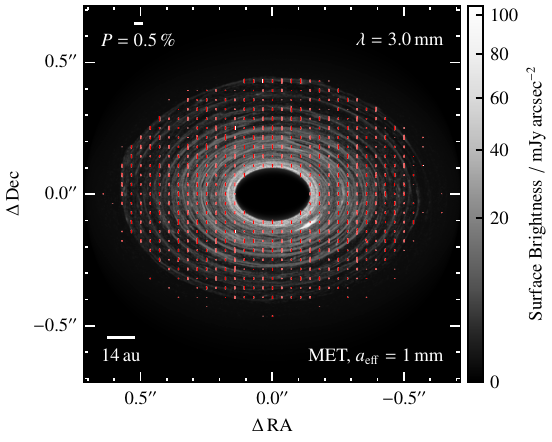}
    \includegraphics[width=\linewidth]{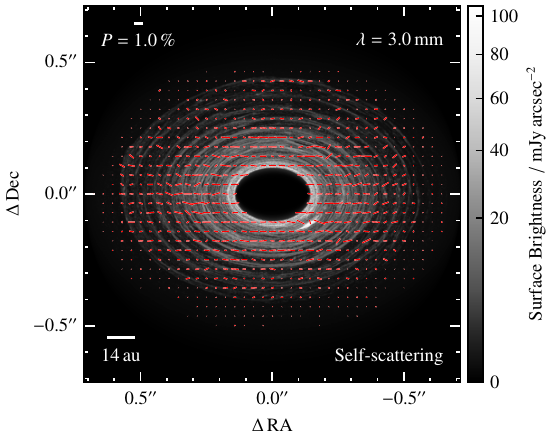}
    \caption{%
        Surface brightness map of the total emitted thermal radiation at a wavelength of \SI{3}{mm} overlaid with corresponding polarization vectors.
        The polarization arises from thermal emission and dichroic absorption of aligned \SI{1}{mm}-sized dust grains (top) and from self-scattered radiation (bottom) only.
        The disk is inclined by \ang{45}.
        See Sect.~\ref{subsec:flip_polarization} for details.
    }
    \label{fig:sb_a1mm_selfsca_i45_w3mm}
\end{figure}

As shown in Fig.~\ref{fig:absorption_eff}, the absorption efficiency along the minor grain axis becomes greater than the absorption efficiency along the major grain axis for particle sizes approximately the wavelength of observation.
As a consequence, the resulting polarization orientation changes.
In particular, for a wavelength of \SI{3}{mm} and aligned grains with a size of \SI{1}{mm}, the polarization orientation of the emitted radiation rotates by an angle of \ang{90} and is parallel to the minor axis of the disk, as shown in Fig.~\ref{fig:sb_a1mm_selfsca_i45_w3mm}.
Here, the polarization vectors coincide with the expected orientation of polarization due to self-scattering, as depicted in Figs.~\ref{fig:sb_align_selfsca_i45}c and~\ref{fig:sb_align_selfsca_i45}d.
In contrast, if only \SI{100}{\um}-sized grains are aligned, the orientation is still parallel to the major axis of the disk for all considered wavelengths.
This is because the flip of the polarization is expected to occur only for wavelengths $\lambda \lesssim 2 \pi a_\mathrm{eff}$.

Most interestingly, the polarization orientation also changes for self-scattered radiation at a wavelength of \SI{3}{mm} (see Fig.~\ref{fig:sb_a1mm_selfsca_i45_w3mm}).
This polarization reversal \citep{Yang_etal_2016} is also caused by dust grains, which are about the size of the observing wavelength.
In particular, the polarization of scattered radiation is parallel to the scattering plane, causing rotation by an angle of \ang{90} in the polarization.
However, the degree of linear polarization of thermally emitted radiation is similar or somewhat lower compared to the polarization due to self-scattering, that is, $\lesssim$\,\SI{1}{\percent}.
Therefore, if both mechanisms are considered at the same time, contributions of polarized emission and self-scattering may cancel each other out.

\subsection{Comparison to an analytic drift model}
\label{subsec:comparison_drift_model}

\begin{figure}
    \centering
    \includegraphics[width=0.991\linewidth]{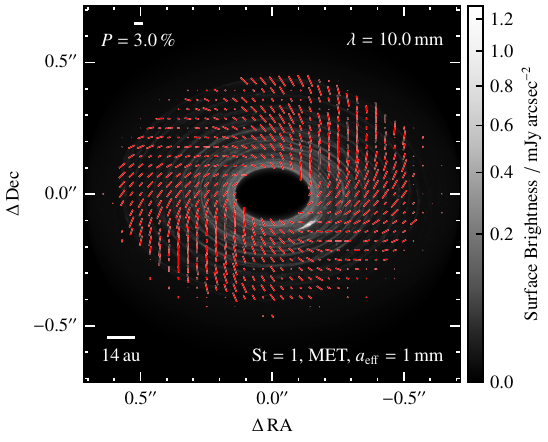}
    \includegraphics[width=0.991\linewidth]{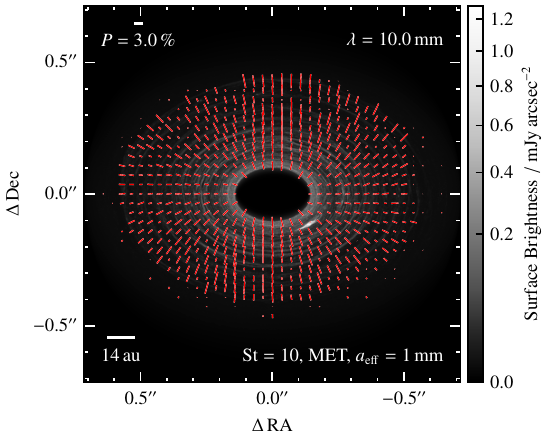}
    \caption{%
        Surface brightness map of the total emitted thermal radiation at a wavelength of \SI{10}{mm} overlaid by polarization vectors assuming an analytic description for the drift velocity (see Eq.~\ref{eq:drift_model}).
        The Stokes number is set to $\mathrm{St} = 1$ (top) and $\mathrm{St} = 10$ (bottom).
        The inclination of the disk is \ang{45}.
        Here, only \SI{1}{mm}-sized grains are aligned.
        See Sect.~\ref{subsec:comparison_drift_model} for details.}
    \label{fig:sb_model_a1mm_i45_w10mm}
\end{figure}

Finally, we compare our results with those expected in the case of an analytic drift model.
In contrast to numerical simulations, here the radial, azimuthal, and polar components of the headwind or drift velocity can be written as%
\begin{equation}
    \label{eq:drift_model}
    \varv_r = \frac{\mathrm{St}}{1 + \mathrm{St}^2} \eta \varv_\mathrm{K}\, , \quad
    \varv_\phi = -\frac{\mathrm{St}^2}{1 + \mathrm{St}^2} \eta \varv_\mathrm{K}\, , \quad
    \varv_\theta \equiv \SI{0}{m.s^{-1}}\, ,
\end{equation}
respectively, where $\mathrm{St}$ is the Stokes number and $\eta \varv_\mathrm{K}$ is the gas rotation velocity relative to the Keplerian velocity \citep{Kataoka_etal_2019}.
Here, we assume $\eta \varv_\mathrm{K} = \SI{53}{m.s^{-1}}$.
In this theoretical drift model, instabilities and turbulences such as the VSI are ignored.
However, the orientation of the drift velocity vector depends on the Stokes number, which causes the radial or azimuthal component to dominate \citep[see][]{Kataoka_etal_2019}.

For the given total disk gas mass (see Table~\ref{tab:model_setup}), Stokes numbers would be in the order of $\lesssim$\,\num{0.1} for the largest \SI{1}{mm}-sized dust grains.
However, for these values, Eq.~\eqref{eq:rayleigh_reduction_factor} is not satisfied.
Thus, the grains are not aligned because of METs and the thermal radiation is unpolarized.
Conversely, this means that with our disk model, a toroidal polarization pattern can only be produced if our initial assumptions for alignment (see Sect.~\ref{subsec:grain_alignment}) are not fulfilled.
In this case, grain alignment due to RATs is an alternative explanation for the azimuthal orientation of polarization \citep{Tazaki_etal_2017}.
In order to present exemplary results of polarization maps, we apply larger Stokes numbers of 1 and 10.
For a Stokes number of 1, the polarization pattern is defined by a transition between a radial and azimuthal orientation.
In contrast, a Stokes number of 10 causes a purely radial polarization pattern.

Figure~\ref{fig:sb_model_a1mm_i45_w10mm} shows the surface brightness map of the emitted thermal radiation at a wavelength of \SI{10}{mm}.
The polarization is caused by elongated \SI{1}{mm}-sized grains that are aligned to the drift velocity vector defined by Eq.~\eqref{eq:drift_model}.

The main difference from the results discussed in Sect.~\ref{subsec:polarization_alignment} is the orientation of the polarization.
In particular, the analytic drift model assumes only radial and azimuthal drift velocity components, whereas the strength of each component depends on the Stokes number.
This is in contrast to the results obtained from hydrodynamical simulations, since the VSI causes a dominating vertical drift velocity component that produces no radial or azimuthal polarization structure.

Moreover, the degree of linear polarization is significantly higher if the alignment is described by the analytic drift velocity model (Eq.~\ref{eq:drift_model}) compared to the polarization if the alignment is described by the velocity field obtained with the PLUTO code.
This is due to the fact that the analytic velocity field described by Eq.~\eqref{eq:drift_model} is homogeneous throughout the model space.
In contrast, the velocity field from numerical hydrodynamical simulations is irregular with disturbances on small scales, causing a decrease in the net degree of polarization.

\subsection{Comparison to observations}
\label{subsec:comparison_observations}

Do we observe polarization due to aligned grains?
With our 3D radiation hydrodynamical models, we obtain a detailed and rich dataset of the kinematic structure of gas and dust of a VSI turbulent protoplanetary disk.
For a VSI turbulent disk, the drift velocity or headwind of the dust grains is dominated by the large-scale vertical motions characteristic of the VSI.
We found that in this model, if the grains are mechanically aligned, the polarization patters are mostly parallel to the major axis of the disk for a large range of millimeter wavelengths (see Figs.~\ref{fig:sb_align_selfsca_i45}a and~\ref{fig:sb_align_selfsca_i45}b).

However, in recent observations of the polarized emission of several protoplanetary disks at ALMA Band~7 \citep{Stephens_etal_2017, Harrison_etal_2024, Lin_etal_2024}, the polarization vectors are parallel to the projected minor axis of the protoplanetary disk, which is more similar to the self-scattering pattern (see Figs.~\ref{fig:sb_align_selfsca_i45}c and~\ref{fig:sb_align_selfsca_i45}d).
At longer wavelengths, in particular at ALMA Band~3, the grains are aligned with their long axis along the azimuthal direction of the disk \citep{Kataoka_etal_2017, Stephens_etal_2017}, probably caused by RATs and not the mechanism considered in this work.
Finally, for the disk around L1448~IRS3B, \citet{Looney_etal_2025} observed polarization that is consistent with the mechanical alignment of dust grains along the spiral dust structures.

More observations of further less massive disks and different instability simulations \citep{Lesur_etal_2023} are needed to refine our understanding of how grains are aligned and how they emit.
We also want to remind the reader that smaller grains that are less settled might shield the emission of larger grains \citep{Sierra_Lizano_2020} and thus dominate the self-scatter emission.
This could be the case of \SI{100}{\um}-sized grains that dominate the self-scatter emission, shielding the emission of millimeter-sized grains drifting at the midplane.


\section{Model assumptions and limitations}
\label{sec:limitations}

There are several limitations in our modeling of the mechanical alignment of large BAM dust aggregates that may affect the synthetic observations presented in this paper.
First, the Monte Carlo nature of the MAD simulations introduces statistical noise in the angular velocity and alignment angle distributions.\footnote{An estimate of the Monte Carlo noise in MAD simulations is provided in Appendix B of \citet{Reissl_etal_2023}.}
We use 30 grain realizations per size bin, which we consider sufficient to capture the mean alignment behavior.
This is because the Rayleigh reduction factor $\mathcal{R}$ is computed using each of the 30 individual Monte Carlo realizations of $\Theta$ (see Eq.~\ref{eq:RRF}).
As a result, $\mathcal{R}$ is an averaged quantity that smooths out both individual outliers in $\Theta$ and the expected Monte Carlo noise.
Increasing the number of realizations would reduce the influence of this noise and could be explored in future studies that focus on the statistical properties of alignment.

Second, although our alignment simulations explicitly cover grain sizes up to \SI{1.4}{\um}, we extrapolate the derived alignment parameters to grains up to \SI{1}{mm} in size, assuming that alignment behavior converges for larger grains (see Appendix~\ref{app:timescales}).
This assumption is motivated by the competing effects of increased surface area and a more spherical morphology, which tend to flatten the torque response across larger sizes.
Although our approach follows the trends reported in \citet{Reissl_etal_2023}, this extrapolation remains an assumption and should be tested directly in future work through high-resolution MAD simulations of larger aggregates.

Finally, our radiative transfer modeling is based on a single hydrodynamical snapshot taken after 750 orbits of disk evolution.
This choice is justified by the assumption that the alignment timescale is shorter than the local dynamical timescales.
However, a more detailed comparison between the alignment timescale and the temporal variability of the disk environment would strengthen this justification.
In principle, time-averaged snapshots may offer a more robust basis for post-processing if temporal variability proves significant.
Although our approach is consistent with previous studies, a more comprehensive, time-resolved treatment of grain alignment may be desirable in future investigations.

We emphasize once again that our modeling of dust polarization focuses exclusively on two distinct effects: (i) dust scattering and (ii) polarized thermal emission from mechanically aligned grains.
However, it is highly unlikely that dust polarization in protoplanetary disks arises from one single mechanism in isolation.
Multiple alignment and polarization mechanisms have been proposed over the years and are expected to operate simultaneously within disk environments \citep[see, e.g.,][]{Andersson_2015}.
Consequently, a comprehensive model of the dust polarization process would require incorporating all relevant mechanisms into a unified RT framework \citep{Hoang_Lazarian_2016, Giang_etal_2023}.
However, the question of how all these mechanisms work in tandem is still a matter of debate.
Hence, such an endeavor is beyond the scope of the present study and will be addressed in future work.


\section{Summary and conclusions}
\label{sec:conclusions}

The goal of this study is to evaluate the impact of mechanical grain alignment in protoplanetary disks on the potentially observable polarization.
The grain alignment is assumed to be the result of the drift motion of the grains.
For this purpose, we combined radiation hydrodynamical simulations to determine the gas and dust density distribution as well as their velocity fields, Monte Carlo dust-gas interaction simulations to calculate the mechanical alignment of dust in a gas flow, and finally Monte Carlo polarized radiative transfer simulations to obtain synthetic observations for the flux and its degree of polarization of thermally re-emitted radiation of aligned nonspherical grains as well as radiation scattered by dust grains.

First, we applied the code PLUTO, which solves the radiation hydrodynamic equations and provides the density distribution (Fig.~\ref{fig:dust_mass_dens}) and the velocity components (see Fig.~\ref{fig:drift_velocity}).
Based on these quantities, the headwind or drift velocity was calculated (see Eq.~\ref{eq:drift_velocity}), which defines the direction of grain alignment due to mechanical alignment.
Subsequently, by applying ballistic aggregation and migration of the primary particle, dust grain aggregates were created (see Sect.~\ref{subsec:grain_aggregation}).
Next, Monte Carlo dust-gas interaction simulations were used to determine a lower and upper grain size limit for alignment in a gas flow (see Sect.~\ref{subsec:grain_alignment}).
Moreover, we estimated the effect of depolarization due to grain precession via the Rayleigh reduction factor (see Eq.~\ref{eq:rayleigh_reduction_factor}).
Finally, POLARIS calculated the temperature of dust grains in the protoplanetary disk and then the thermal and self-scattered radiation by applying Monte Carlo radiative transfer simulations (see Sect.~\ref{subsec:radiative_transfer}).
For the calculation of polarization due to dichroic emission and absorption, nonspherical grains were represented by oblate shaped particles that are aligned with their minor axis to the drift velocity vector.
The optical properties of these dust grains were calculated using DDSCAT, whereas for temperature and scattering simulations, the optical properties were calculated using the Mie scattering theory (see Sect.~\ref{subsec:grain_properties}).

We find that large dust grains, which contribute the most to the observable polarization, fulfill the alignment criteria (see Eq.~\ref{eq:rayleigh_reduction_factor} and Fig.~\ref{fig:thermal_vel_xy}) and are therefore aligned due to the gas flow.
In particular, because of hydrodynamical instabilities such as the VSI, the polar drift velocity that is perpendicular to the disk midplane is the dominant velocity component (see Fig.~\ref{fig:drift_velocity}).
As a result, most grains align with their short axis parallel to the rotational axis of the disk.
Consequently, the orientation of the polarization of the thermal emission is close to zero for a face-on disk, but parallel to the major disk axis for an inclined disk (see Fig.~\ref{fig:sb_align_selfsca_i45}).
Here, the highest degree of polarization is found for the longest considered wavelength of \SI{10}{mm}.
Most notably, the orientation of polarization is in contrast to analytic drift velocity models, which assume solely radial and azimuthal drift components (see Sect.~\ref{subsec:comparison_drift_model}).

In addition, polarization arises from self-scattering as well (see Sect.~\ref{subsec:polarization_scattering}).
In this case, the orientation of the polarization is usually parallel to the projected minor disk axis.
Here, the contribution of self-scattering is largest for the shortest considered wavelength of \SI{0.87}{mm}.
With increasing wavelength, and therefore decreasing optical depth, the impact of self-scattering becomes less important in the outer disk regions.

Moreover, the VSI potentially forms vortices that can trap particles.
Due to the increase in dust density, the surface brightness increases near the vortex as well (see Sect.~\ref{subsec:polarization_vortex}).
In contrast, the degree of polarization from thermally emitted radiation decreases, whereas the polarization from self-scattering produces a nonuniform pattern because of the anisotropic radiation field.

Finally, we observe a change in the orientation of the polarization by an angle of \ang{90}.
This behavior is expected if the observing wavelength is about the size of the dust grains, that is, $\lambda \lesssim 2 \pi a$.
In particular, at a wavelength of \SI{3}{mm}, the orientation changes in the polarization of thermal emission of aligned \SI{1}{mm}-sized dust grains as well as in the polarization due to self-scattering (see Fig.~\ref{fig:sb_a1mm_selfsca_i45_w3mm}).
Most interestingly, the polarization vectors for the thermally emitted radiation are parallel to the projected minor disk axis and coincide with the expected case of polarization due to self-scattering, as depicted in Fig.~\ref{fig:sb_align_selfsca_i45}.

In conclusion, grain alignment driven solely by mechanical torques proves to be an efficient mechanism for aligning millimeter-sized grains in protoplanetary disks, particularly when hydrodynamical instabilities such as the VSI are taken into account.
Our synthetic polarimetric observations in the millimeter wavelength range demonstrate that the resulting polarization may serve as a promising diagnostic tool to trace the orientation of aligned millimeter-sized dust grains.


\begin{acknowledgements}
    The authors thank the anonymous referee for useful and constructive comments.
    M.L.-S. and S.W. acknowledge financial support from the DLR/BMWK grant 50 OR 2210.
    S.R. acknowledge financial support from the Heidelberg cluster of excellence (EXC 2181 -- 390900948) “\emph{STRUCTURES}: A unifying approach to emergent phenomena in the physical world, mathematics, and complex data”, specifically via the exploratory project EP 4.4.
    Furthermore, S.R. thanks for funding from the European Research Council in the ERC synergy grant “\emph{ECOGAL} -- Understanding our Galactic ecosystem: From the disk of the Milky Way to the formation sites of stars and planets” (project ID 855130).
    This research made use of
    \href{https://github.com/astropy/astropy}{Astropy} \citep{Astropy_etal_2013, Astropy_etal_2018, Astropy_etal_2022},
    \href{https://github.com/matplotlib/matplotlib}{Matplotlib} \citep{Hunter_2007},
    \href{https://github.com/numpy/numpy}{NumPy} \citep{Harris_etal_2020},
    \href{https://github.com/scikit-image/scikit-image}{scikit-image} \citep{van_der_walt_etal_2014},
    and the \href{https://ui.adsabs.harvard.edu}{Astrophysics Data System}, funded by NASA under Cooperative Agreement 80NSSC21M00561.
\end{acknowledgements}

\bibliographystyle{aalink}
\bibliography{bibliography.bib}


\begin{appendix}

\section{Characteristic alignment timescale}
\label{app:timescales}

In this section, we briefly evaluate the extrapolation of the alignment process for large grains, that is, aggregates with an effective radius of \SI{100}{\um} and \SI{1}{mm}, respectively.

The time-dependent magnitude of the angular velocity, $\omega(t)$, of each grain can be evaluated via
\begin{equation}
\omega(t)= \omega_{\mathrm{max}}\left[1-\exp\left( - \frac{t}{\tau_{\mathrm{drag}}} \right) \right]
\end{equation}
by solving the differential equation Eq.~\eqref{eq:time_evolution}.
For simplicity, we use the maximum mechanical torque $\Gamma_{\mathrm{MET}} (\Theta)$ over all angles $\Theta$ to estimate the upper limit of the maximum angular momentum, ${ \omega_{\mathrm{max}} = \max_\Theta \left(\Gamma_{\mathrm{MET}} (\Theta) \right)\tau_{\mathrm{drag}} / I_{\mathrm{m}} }$, of each individual BAM1 dust aggregate.
Consequently, each aggregate reaches its maximum angular momentum, $ \omega_{\mathrm{max}}$, within a \SI{99}{\percent} limit over a characteristic drag timescale of ${ 5\times \tau_{\mathrm{drag}} }$.

\begin{figure*}
    \centering
    \includegraphics[width=0.98\linewidth]{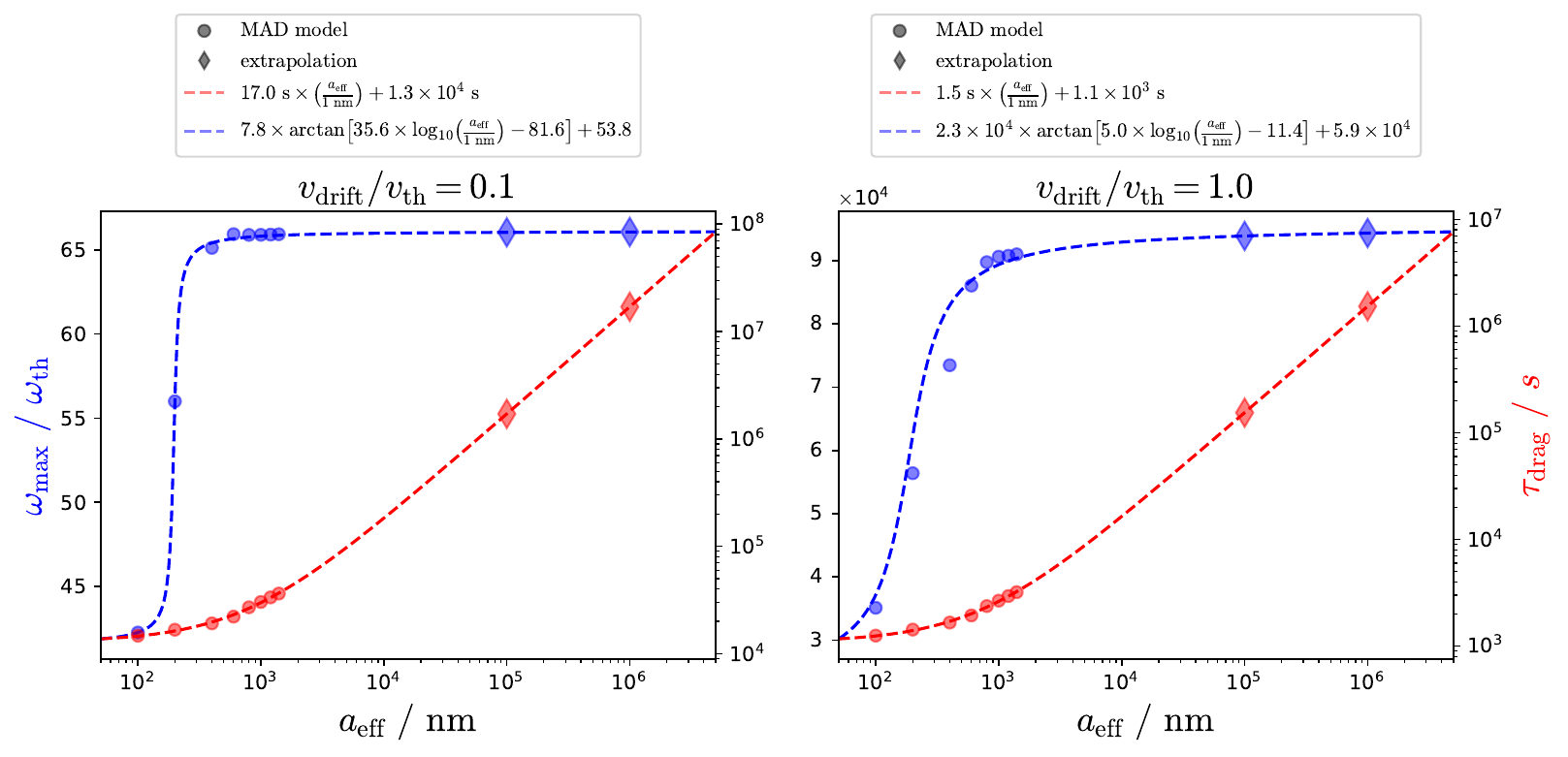}
    \caption{
        Average alignment over all BAM1 grain shapes according to the MAD model (dot shapes) and their extrapolation (diamond shapes) for the drag timescale $t_{\mathrm{drag}}$ (red) and the ratio of angular velocity $\omega$ (blue) with respect to the thermal angular velocity $\omega_{\mathrm{th}}$ dependent on the effective grain radius $a_{\mathrm{eff}}$.
        The data is modeled for the fiducial values of $n_{\mathrm{g}} = \SI{e10}{cm^{-3}}$, $T_{\mathrm{g}} = \SI{20}{K}$, and $T_{\mathrm{d}} = \SI{20}{K}$.
        The left panel depicts data for the distinct ratio of the drift velocity $\varv_{\mathrm{drift}}$ to the thermal velocity $\varv_{\mathrm{th}}$ with $\varv_{\mathrm{drift}} / \varv_{\mathrm{th}} = 0.1$ while the right panel if for $\varv_{\mathrm{drift}} / \varv_{\mathrm{th}} = 1.0$.
        The best fits (dashed lines) of $\tau_{\mathrm{drag}}$ are performed via ${ \tau_{\mathrm{drag}} = a \times \left( a_{\mathrm{eff}} / \SI{1}{nm} \right) + b}$ and ${ \omega / \omega_{\mathrm{th}} = a \times \arctan\left[b \times \log_{\mathrm{10}} \left( a_{\mathrm{eff}} / \SI{1}{nm} \right) + c \right] + d}$, respectively.
        The MAD data is extrapolated data for the two particular grain species of $a_{\mathrm{eff}} = \SI{100}{\um}$ as well as $a_{\mathrm{eff}} = \SI{1}{mm}$ utilized in this paper.
    }
    \label{fig:params}
\end{figure*}

In Fig.~\ref{fig:params}, we present the ratio of the maximum angular velocity, $\omega_{\mathrm{max}}$, to the thermal angular velocity $\omega_{\mathrm{th}}$, averaged over all ensembles of dust aggregates with the same effective radius $a_{\mathrm{eff}}$, for velocity ratios of $\varv_{\mathrm{drift}} / \varv_{\mathrm{th}} = 0.1$ and $\varv_{\mathrm{drift}} / \varv_{\mathrm{th}} = 1.0$.
Although some ensembles may contain outliers, the ensemble average shows a clear trend in which the ratio ${\omega_{\mathrm{max}} / \omega_{\mathrm{th}} }$ approaches an asymptotic value with increasing $a_{\mathrm{eff}}$.
This asymptotic behavior is consistent with the findings of \cite{Reissl_etal_2023}, where the mechanical spin-up process was modeled using fractal dust aggregates.
However, in \cite{Reissl_etal_2023}, only aggregates with an upper grain radius of $a_{\mathrm{eff}} = \SI{800}{nm}$ were considered, whereas in this publication we use aggregates up to $a_{\mathrm{eff}} = \SI{1.4}{\um}$.
Using aggregates up to \SI{1.4}{\um} reveals the asymptotic trend even more clearly.

We again emphasize that the simulation of mechanical alignment parameters beyond \SI{1.4}{\um} aggregates is numerically unfeasible \citep[see][]{Reissl_etal_2023}.
To quantify the trend, we perform a least-squares fit with an $\arctan$ function.
As shown in Fig.~\ref{fig:params}, the ratio ${\omega_{\mathrm{max}} / \omega_{\mathrm{th}} }$ already approximates its asymptote for aggregates with $a_{\mathrm{eff}} \geq \SI{1.0}{\um}$.
Hence, in this paper, we assume that grains with $a_{\mathrm{eff}} = \SI{100}{\um}$ and \SI{1}{mm} have a similar ratio of ${\omega_{\mathrm{max}} / \omega_{\mathrm{th}} }$ as the much smaller aggregates within the range $a_{\mathrm{eff}} = \SI{1.0}{\um}$ to \SI{1.4}{\um}.

We extrapolate the characteristic drag timescale, $\tau_{\mathrm{drag}}$, for the considered aggregate sizes of \SI{100}{\um} and \SI{1}{mm} similarly.
Since the drag depends only marginally on IR photon emission due to the low dust temperature within the disk, it is dominated by the gas drag timescale $\tau_{\mathrm{gas}} \propto I_{\mathrm{m}} n_{\mathrm{g}}^{-1} a_{\mathrm{eff}}^{-4}$ \citep[see e.g.][]{Draine_Weingartner_1996,Jaeger_etal_2024}.
Given that the maximal moment of inertia of each aggregate scales with $I_{\mathrm{m}} \propto a_{\mathrm{eff}}^{5}$, we obtain a nearly linear dependence of $\tau_{\mathrm{drag}} \propto a_{\mathrm{eff}}$.

In Fig.~\ref{fig:params}, we show $\tau_{\mathrm{drag}}$ as a function of $a_{\mathrm{eff}}$ for the ratios ${ \varv_{\mathrm{drift}} / \varv_{\mathrm{th}} = 0.1 }$ and ${ \varv_{\mathrm{drift}} / \varv_{\mathrm{th}} = 1.0 }$, respectively.
Based on the average $\tau_{\mathrm{drag}}$ across all aggregate ensembles of similar size, we linearly extrapolate $\tau_{\mathrm{drag}}$ up to the utilized grain radii of $a_{\mathrm{eff}} = \SI{100}{\um}$ and \SI{1}{mm}, respectively.
For a ratio of $\varv_{\mathrm{drift}} / \varv_{\mathrm{th}} = 0.1$, the large aggregates exhibit drag timescales of \SI{1.71e6}{s} and \SI{1.70e7}{s}.%
\footnote{These particular values of the gas drag timescale $\tau_{\mathrm{drag}}$ are still comparable to those calculated with Eq.~13 presented in \cite{Draine_Weingartner_1996} assuming spherical grains and no drift velocity $\varv_{\mathrm{drift}}$ at all whereas higher $\varv_{\mathrm{drift}}$ would decrease $\tau_{\mathrm{drag}}$.}
In contrast, for $\varv_{\mathrm{drift}} / \varv_{\mathrm{th}} = 1.0$, the drag timescales decrease to \SI{1.54e5}{s} and \SI{1.53e6}{s}, respectively, assuming a gas number density of $n_{\mathrm{g}} = \SI{e10}{cm^{-3}}$ and a gas temperature of $T_{\mathrm{g}} = \SI{20}{K}$.
Hence, the largest considered grain size with $a_{\mathrm{eff}} = \SI{1}{mm}$ would reach its final grain alignment configuration, that is, an angular velocity $\omega_{\mathrm{max}}$, within ${ 5\times \tau_{\mathrm{drag}} \approx \SI{3}{yr} }$ at most based on the extrapolated data from MAD simulations.
For increasing gas density, these timescales become even shorter.
We emphasize that micro-turbulence of the gas on aggregate scales is already accounted for in the MAD model, where the magnitude and direction of the gas velocity are sampled from a distribution function based on gas temperature and drift velocity.

Therefore, even the largest aggregates reach their final alignment configuration on timescales significantly shorter than that of the vertical shear instability, which operates on the order of tens of orbital timescales, which represents several thousands of years at \SI{50}{au}.

\FloatBarrier


\section{Direction of grain alignment}
\label{app:alignment}

Usually, grain alignment and the resulting dust polarization are estimated using toy models or dust grains with one idealized shape throughout \citep{Purcell_Spitzer_1971, Das_Weingartner_2016, Hoang_etal_2018}.
The advantage of the MAD framework is that it considers the mechanical alignment of a large ensemble of irregularly shaped grains, enabling the evaluation of the expected dust polarization based on the resulting distribution of the alignment angle $\Theta$.

\begin{figure*}
    \centering
    \includegraphics[width=0.92\textwidth]{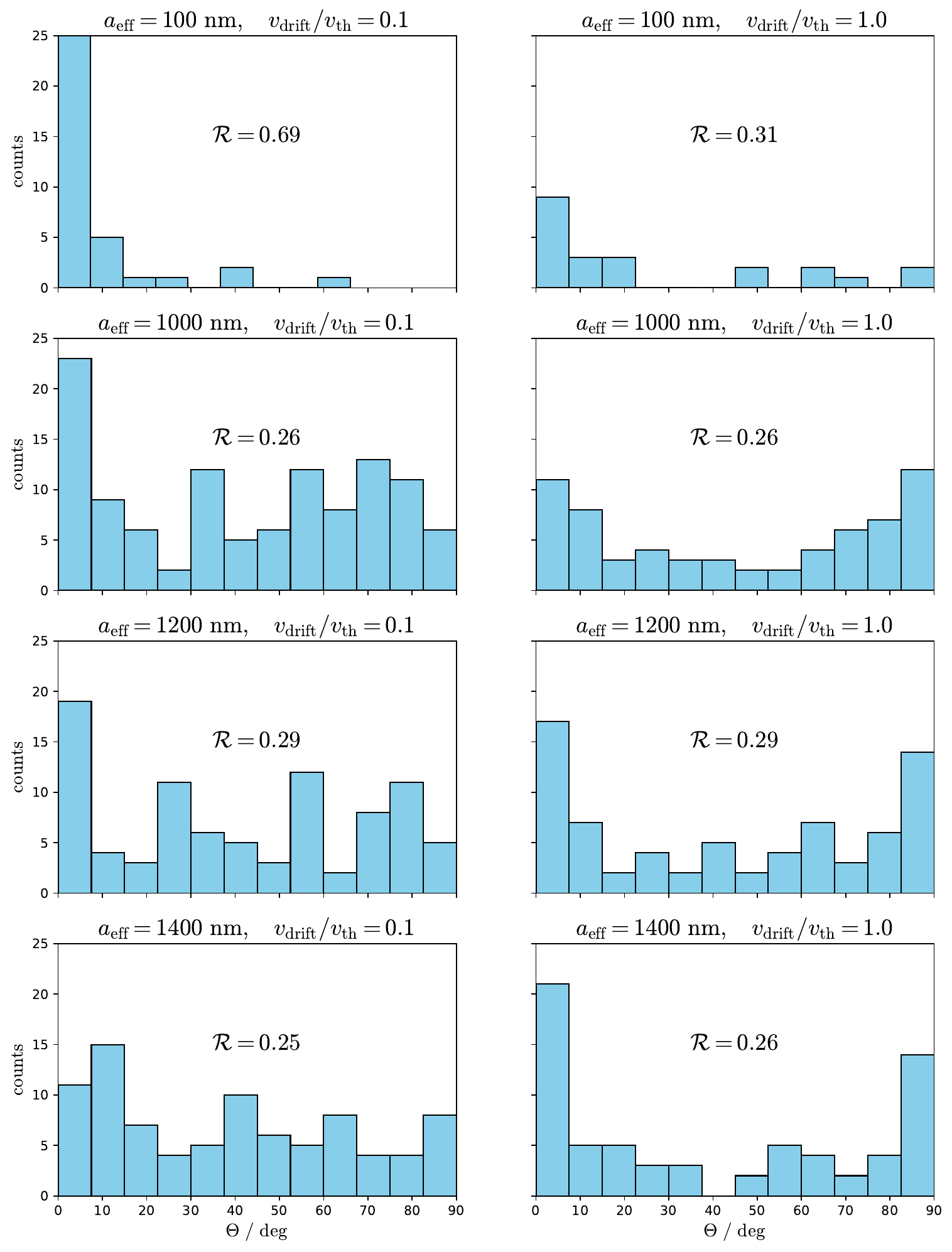}
    \caption{%
        Distribution of the alignment angle $\Theta$ (compare Fig.~\ref{fig:MET_skecth}) of a distinct ensemble of BAM1 dust aggregates according to the MAD framework.
        The four rows (top, second, third, bottom) show the distribution for an effective radius of $a_{\mathrm{eff}} = \SI{100}{nm}$, $a_{\mathrm{eff}} = \SI{1000}{nm}$, $a_{\mathrm{eff}} = \SI{1200}{nm}$, and $a_{\mathrm{eff}} = \SI{1400}{nm}$, respectively.
        Individual alignment angles $\Theta$ are simulated for fiducial values of $n_{\mathrm{g}} = \SI{e10}{cm^{-3}}$, $T_{\mathrm{g}} = \SI{20}{K}$, and $T_{\mathrm{d}} = \SI{20}{K}$ whereas the left column is for the ratio of velocities $\varv_{\mathrm{drift}} / \varv_{\mathrm{th}} = 0.1$ while the right column is for the ratio $\varv_{\mathrm{drift}} / \varv_{\mathrm{th}} = 1.0$.
        The Rayleigh reduction factor $\mathcal{R}$ (see Eq.~\ref{eq:RRF}) is calculated considering both the distribution of $\Theta$ within \ang{0} to \ang{90} and the ratio $\zeta$ of grains that either do not align with the velocity field at all because of slow rotation or grains that are rotationally destroyed.
    }
    \label{fig:theta_dist}
\end{figure*}

In Fig.~\ref{fig:theta_dist} we present the distribution of $\Theta$ for selected grain ensembles with a distinct effective radius between $a_{\mathrm{eff}} = \SI{100}{nm}$ and $\SI{1.4}{\um}$ dependent on the two different velocity ratios of ${ \varv_{\mathrm{drift}} / \varv_{\mathrm{th}} = 0.1 }$ and ${ \varv_{\mathrm{drift}} / \varv_{\mathrm{th}} = 1.0 }$.
For the ratio ${ \varv_{\mathrm{drift}} / \varv_{\mathrm{th}} = 0.1 }$ and grains with $a_{\mathrm{eff}} = \SI{100}{nm}$, the distribution is very narrow close to $\Theta = \ang{0}$.
Hence, the resulting Rayleigh reduction factor amounts to $\mathcal{R} = 0.692$.
For a ratio of ${ \varv_{\mathrm{drift}} / \varv_{\mathrm{th}} = 1.0 }$, the Rayleigh reduction factor decreases to $\mathcal{R} = 0.311$.
In this particular case, the lower value of $\mathcal{R}$ is because part of the aggregates within the ensemble becomes rotationally disrupted (compare Fig.~\ref{fig:MET_alignment}).
Larger aggregates with $a_{\mathrm{eff}} \geq \SI{1}{\um}$ rotate slower in our model because they are more spherically shaped and cannot be rotationally destroyed for ${ \varv_{\mathrm{drift}} / \varv_{\mathrm{th}} \lesssim 10 }$.
However, as shown in Fig.~\ref{fig:theta_dist} the distribution of $\Theta$ for aggregates with $a_{\mathrm{eff}} \geq \SI{1}{\um}$ becomes even broader compared to smaller grains with a second peak appearing closer to $\Theta = \ang{90}$.
Here, the resulting Rayleigh reduction factor is even lower with an average value of $\mathcal{R} = 0.29$.
We report similar distributions of $\Theta$ for the entire range ${ \num{1.78e-2} \lesssim \varv_{\mathrm{drift}} / \varv_{\mathrm{th}} \lesssim 10 }$.

Although we observe convergence in the Rayleigh reduction factor, $\mathcal{R}$, for grain sizes larger than approximately \SI{1}{\um}, we note that individual angle distributions can exhibit bimodal features, particularly under near-sonic drift conditions, as shown in Fig.~\ref{fig:theta_dist}.
These features resemble the classical Gold-type alignment, in which grains align with their short axis perpendicular to the drift velocity \citep{Gold_1952a, Gold_1952b}.
However, our results show a much broader distribution across the full range ${ \ang{0} < \Theta < \ang{90} }$, including a peak near $\Theta = \ang{0}$, which is not captured by the classical Gold mechanism.
In this regard, the MAD framework provides an improvement by accounting for more realistic, irregular grain shapes and their complex gas-grain interactions, even compared to more recent mechanical alignment models \citep[see, e.g.,][]{Lazarian_Hoang_2007b, Das_Weingartner_2016, Hoang_etal_2018}. Moreover, the drift velocities in the utilized MHD snapshot are around 0.1 to 0.2 Mach, that is, subsonic i.e. Gold alignment would not predict any dust polarization at all.

In summary, we see a convergence of angular momentum for larger grains (see Figs.~\ref{fig:MET_alignment} and~\ref{fig:params}) and characteristic grain alignment timescales that are significantly shorter than those of the VSI.
Consequently, for the utilized grain sizes of \SI{100}{\um} and \SI{1}{mm}, we extrapolate a Rayleigh reduction factor of $\mathcal{R} \approx 0.29$ within the velocity ratio range ${ \num{1.78e-2} \lesssim \varv_{\mathrm{drift}} / \varv_{\mathrm{th}} \lesssim 10 }$.

\onecolumn

\section{Velocity field}
\label{app:velocity_field}

\begin{figure*}[!h]
    \centering
    \includegraphics[width=\linewidth]{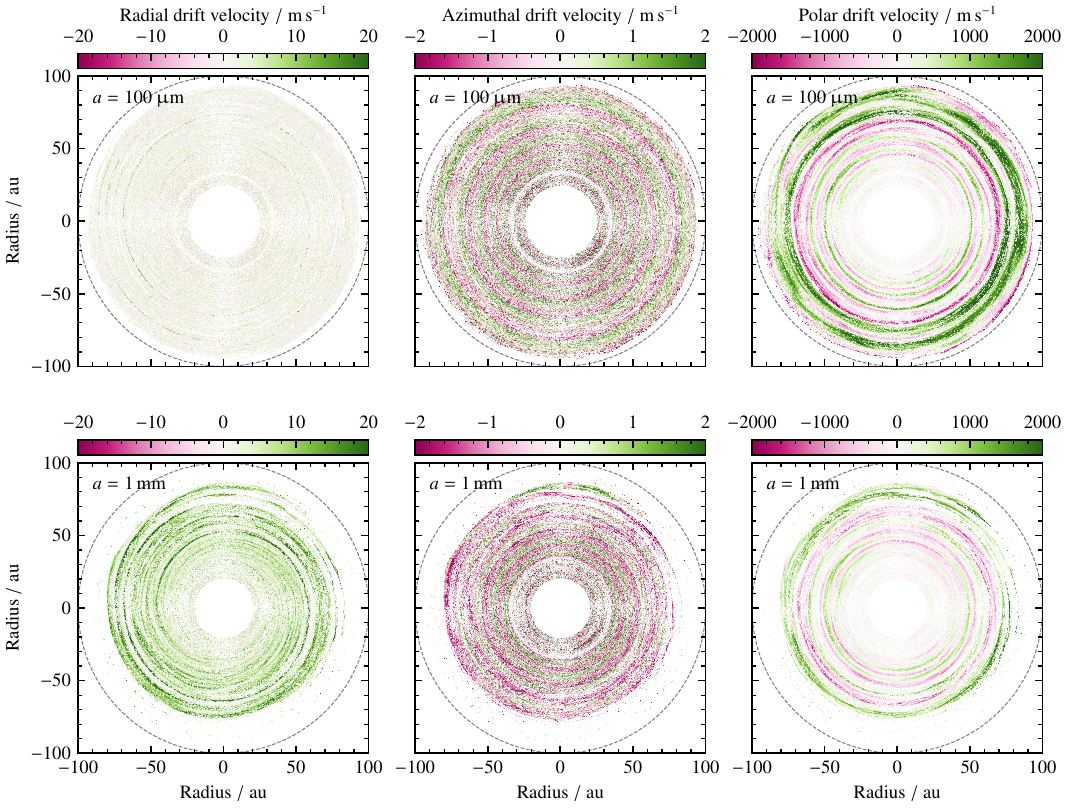}
    \caption{%
        Radial (left), azimuthal (middle), and polar (right) drift velocity component in the disk midplane of \SI{100}{\um} (top) and \SI{1}{mm}-sized (bottom) grains.
        Note the different scales of the color bar for each component of the drift velocity.
        The gray dashed line indicates the radius of the disk at \SI{100}{au}.
    }
    \label{fig:drift_velocity}
\end{figure*}

Figure~\ref{fig:drift_velocity} shows the three components of the drift velocity vector in the disk midplane.
Here, the polar drift velocity is the dominant component, causing alignment with the short axis of the dust grains perpendicular to the disk midplane.
The alternating polar velocity of the \SI{1}{mm}-sized grains at the inner disk rim is a numerical effect caused by the boundary conditions.

\twocolumn

\section{Dust temperature distribution}
\label{app:temperature_distribution}

\begin{figure}[!h]
    \centering
    \includegraphics[width=\linewidth]{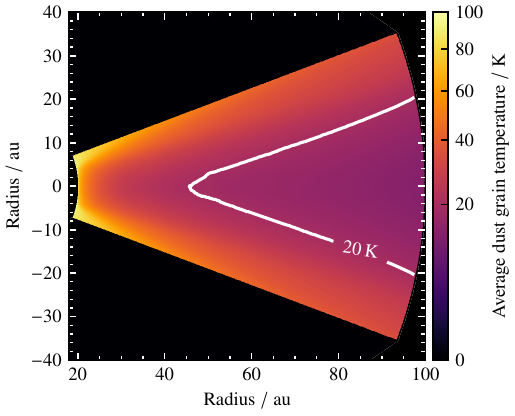}
    \caption{%
        Cross-section view of the average dust grain temperature.
        The temperature values are averaged azimuthally.
    }
    \label{fig:avg_dust_temp}
\end{figure}

Figure~\ref{fig:avg_dust_temp} shows the average dust grain temperature of the circumstellar disk model in a cross-section view resulting from a simulation using POLARIS.
The temperature values are averaged azimuthally.
At the inner rim of the disk, the temperatures are up to about \SI{100}{K}.

\end{appendix}

\end{document}